\documentclass{emulateapj}

\usepackage{amsmath}
\usepackage{amssymb}
\usepackage{graphicx}
\usepackage{epsfig}
\usepackage{color}
\usepackage{natbib}
\newcommand{\Ms}{{\ensuremath{{M}_{\odot} }}}
\newcommand{\Zs}{\ensuremath{Z_\odot}}
\newcommand{\HII}{{\ion{H}{2}}}

\shorttitle{GRBs of the First Stars}
\shortauthors{Mesler et al.}

\begin{document}
\doublespace

\title{The First Gamma-Ray Bursts in the Universe}

\author{R. A. Mesler\altaffilmark{1}, Daniel J. Whalen\altaffilmark{2,3}, Joseph 
Smidt\altaffilmark{2}, Chris L. Fryer\altaffilmark{4}, N. M. 
Lloyd-Ronning\altaffilmark{4} and Y. M. Pihlstr\"om\altaffilmark{1}}

\altaffiltext{1}{Department of Physics and Astronomy, University of New Mexico, 
Albuquerque, NM  87131}
\altaffiltext{2}{T-2, Los Alamos National Laboratory, Los Alamos, NM 87545}
\altaffiltext{3}{Universit\"{a}t Heidelberg, Zentrum f\"{u}r Astronomie, Institut f\"{u}r 
Theoretische Astrophysik, Albert-Ueberle-Str. 2, 69120 Heidelberg, Germany}
\altaffiltext{4}{CCS-2, Los Alamos National Laboratory, Los Alamos, NM 87545}
 
\begin{abstract}

Gamma-ray bursts (GRBs) are the ultimate cosmic lighthouses, capable of 
illuminating the universe at its earliest epochs.  Could such events probe the 
properties of the first stars at $z \sim$ 20, the end of the cosmic Dark Ages?  
Previous studies of Population III GRBs only considered explosions in the 
diffuse relic \HII\ regions of their progenitors, or bursts that are far more more 
energetic than those observed to date. But the processes that produce GRBs 
at the highest redshifts likely reset their local environments, creating much 
more complicated structures than those in which relativistic jets have been 
modeled so far.  These structures can greatly affect the luminosity of the 
afterglow, and hence the redshift at which it can be detected.  We have now 
simulated Population III GRB afterglows in \HII\ regions, winds, and dense 
shells ejected by the star during the processes that produce the burst.  Our 
model, which has been used in previous work, has been extended to include 
contributions from reverse shocks, inverse Compton cooling and the effects of 
sphericity and beaming in the blast wave, and is valid in a variety of circumjet 
density profiles.  We find that GRBs with $E_{\mathrm{iso},\gamma} =$ 10$^{
51}$ - 10$^{53}$ erg will be visible at $z \gtrsim$ 20 to the next generation of 
near infrared and radio observatories.  In many cases, the environment of the 
burst, and hence progenitor type, can be inferred from the afterglow light curve.  
Although some Population III GRBs are visible to {\it Swift} and the Very Large 
Array now, the optimal strategy for their detection will be future missions like 
EXIST and JANUS, which have large survey areas and onboard X-ray and 
infrared telescopes that can track their near infrared flux from the moment of 
the burst, thereby identifying its redshift.

\end{abstract}


\section{Introduction}

Gamma-ray bursts (GRBs) are the ultimate cosmic lighthouses, capable of 
illuminating the universe at its earliest times.  They have now been detected 
at $z =$ 9.4 \citep[GRB 090423;][]{cucc11}, $z =$ 8.26 \citep[GRB 090423;][]{
salv09}, and $z = $ 6.7 \citep[GRB 080913;][]{grein09}.  Besides tracing star 
formation rates over cosmic time \citep[e.g.,][]{tot97,wij98}, GRBs can also 
constrain the metallicity and reionization history of the early IGM \citep[e.g.,][]{
i03,tot06,wang12}, the dark energy equation of state \citep[e.g.,][]{wang11}, 
and the properties of host galaxies \citep{brk04,toma11} \citep[see also][for a 
concise discussion of GRB cosmology]{bl06b}.  

Could GRBs also probe the properties of the first stars, and the environments 
in which they form?  Population III (Pop III) stars are the key to understanding
early cosmological reionization \citep[e.g.,][]{ket04,awb07} and chemical 
enrichment \citep{ss07,bsmith09,ritt12,ss13}, the properties of primeval 
galaxies \citep[e.g.,][]{jeon11,wise12,pmb12} and the origins of supermassive 
black holes \citep[e.g.,][]{wf12,jlj12a,jet13,latif13c}. Although numerical models 
\citep[e.g.,][]{fsg09,glov12,dw12,hir13} and the fossil abundance record \citep{
bc05,fet05,jet09b,jw11,caffau12,keller14} both suggest that Pop III stars are 
30 - 500 \Ms, there are, for now, no direct observational constraints on their 
masses.  

Pop III GRBs might signal the deaths of the first stars because they are visible
at very high redshifts.  Gamma rays from long-duration GRBs can be detected 
by \textit{Swift} out to $z \gtrsim$ 20 \citep{lr00,bl02,bl06a,mr10}.  Analytical 
models suggest that GRB afterglows, which are required to pinpoint their 
redshifts, may be visible at $z \sim$ 15 - 30 \citep{cl00,gou04,im05,inoue07,
ds11,nak12,kash13}.  These studies either did not have complete afterglow 
physics or only examined jets in uniform densities, like those of the \HII\ regions 
of Pop III stars \citep{wan04}\citep[but see][for a $z \sim$ 6 GRB that is thought 
to have occurred in a uniform density]{edo14}.  

But the processes that produce a GRB likely reset its local environment, creating 
far more complicated structures than those in which GRB jets have been 
modeled to date.  These structures, such as multiple shocks and dense shells, 
can greatly affect the luminosity of the afterglow and hence the redshift at which 
it can be detected.  The latest work has focused on the very energetic bursts 
\citep[10$^{55}$ - 10$^{57}$ erg;][]{suwa11,nsi12} of 1000 \Ms\ Pop III stars. 
However, Pop III stars this massive have been rendered much less likely by the 
most recent primordial star formation models, and it is not clear if the proto-black 
holes of such stars have the energy densities required to launch GRB jets, given 
their large radii \citep{fwh01,fh11}.  

To obtain more realistic light curves for Pop III GRBs in the near infrared (NIR), 
X-rays, and radio, and to determine more accurate limits in redshift for their
detection, we have modeled the afterglows of these explosions for a variety of 
progenitors and ambient media.  We also determine if the environment of the 
burst, and thus the properties of its progenitor, can be extracted from its light
curves \citep[e.g.,][]{wet08c,met12a}.  These calculations span the usual 
energies for both merger and single-star events, and our afterglow model 
produces light curves for relativistic jets in any density profile, not just the \HII\ 
regions of earlier studies.  Our simulation code is the one used in \citet{met12a} 
(hereafter M12) but with significant improvements that include the contribution 
to the flux by inverse Compton scattering and reverse shocks, and the effects 
of beaming and the spherical nature of the blast wave. 

We also revisit the question of whether or not collisions of GRB jets with large 
density jumps, like those associated with the massive structures ejected by 
some stars prior to the burst, can produce flares.  Standard afterglow models 
predict flares for some collisions that are consistent with those observed in past 
events.  Until recently, such features have usually been attributed to delayed 
energy injection from the central engine, not collisions.  But some contend that 
collisions cannot produce flares because the extended emission region behind 
the blast wave and effects from reverse shocks tend to wash out bumps in the 
light curve \citep{gatEA13}.  In this paper, we explore this problem in greater 
detail.  As shown later, these effects (which our improved algorithm captures) 
mitigate flares but do not eliminate them, and some Pop III GRBs do exhibit
prominent flares.

In Section 2 we review likely pathways for Pop III GRBs, and in Section 3 we 
examine the circumburst environments they create.  We lay out our grid of 
simulations in Section 4 and in Section 5 we present x-ray, NIR, and radio 
light curves for Pop III GRBs in all of these environments and revisit the 
production of flares in some of these events.  Detection strategies for Pop III 
GRBs are examined in Section 6.  Our GRB light curve model is described in
detail in the Appendix.

\section{Pop III GRB Progenitors}

Long duration (LD) GRBs have been shown to be connected to the deaths of 
massive stars \citep[e.g.,][]{stanekEA03} and to Type Ib/c supernovae in 
particular, whose progenitors have lost their hydrogen envelopes \citep{woo06}.  
The leading contender for the central engines of LD GRBs is the collapsar model 
\citep{woo93,mcf01}, in which the core of a massive star collapses to a black hole 
(BH) accretion disk system that drives a relativistic jet through the outer layers of 
the star and into the surrounding medium.  Besides the ejection of the hydrogen 
envelope, which is usually necessary for the jet to break out of the star, collapsars 
require stellar cores with unusually high angular momenta.  In principle, any star 
that creates a black hole can make a GRB, but given the steep decline in the 
stellar IMF in galaxies today, most GRB progenitors are thought to be 40 - 60 \Ms.

Two primary channels have been proposed for LD GRBs.  In the first, a single 
rapidly rotating star sheds its outer envelope in some type of outburst, like a 
luminous blue variable (LBV) ejection \citep[e.g.,][]{baraffe01}, a Wolf-Rayet 
(WR) phase, or a pulsational pair instability \citep{hw02,wet13d}. In the second, 
the progenitor is in a binary when it becomes a red giant. The two stars enter a 
common envelope phase in which the second star is engulfed by the first and 
slowly spirals into its center, ejecting its outer envelope and spinning up its core 
in an exchange of angular momentum.

About a dozen pathways have been proposed in which a tightly-coupled binary 
system can collapse to form a BH accretion disk that powers a GRB, but they 
generally fall into two categories.  In the first, the binary companion is another 
star \citep{fwh99,pasp07} and in the second it is a BH or neutron star (NS) 
\citep[so-called He mergers;][]{fw98,zf01}.  The key difference between the two 
is the time between the ejection of the envelope and the GRB.  In the first, the 
H layer can be ejected as a dense shell up to several hundred kyr before the 
death of the star and have a radius of several pc at the time of the burst.  In
most cases, this shell will be beyond the reach of the jet. In the second, a slower, 
more massive shell is ejected only a few years before the orbit of the BH decays 
into the center of the star and forms an accretion disk.  The shell may only be a 
few AU in radius at the time of the burst.  In both cases, strong winds usually 
precede and follow the expulsion of the envelope.

\begin{figure*}
\begin{center}
\begin{tabular}{cc}
\epsfig{file=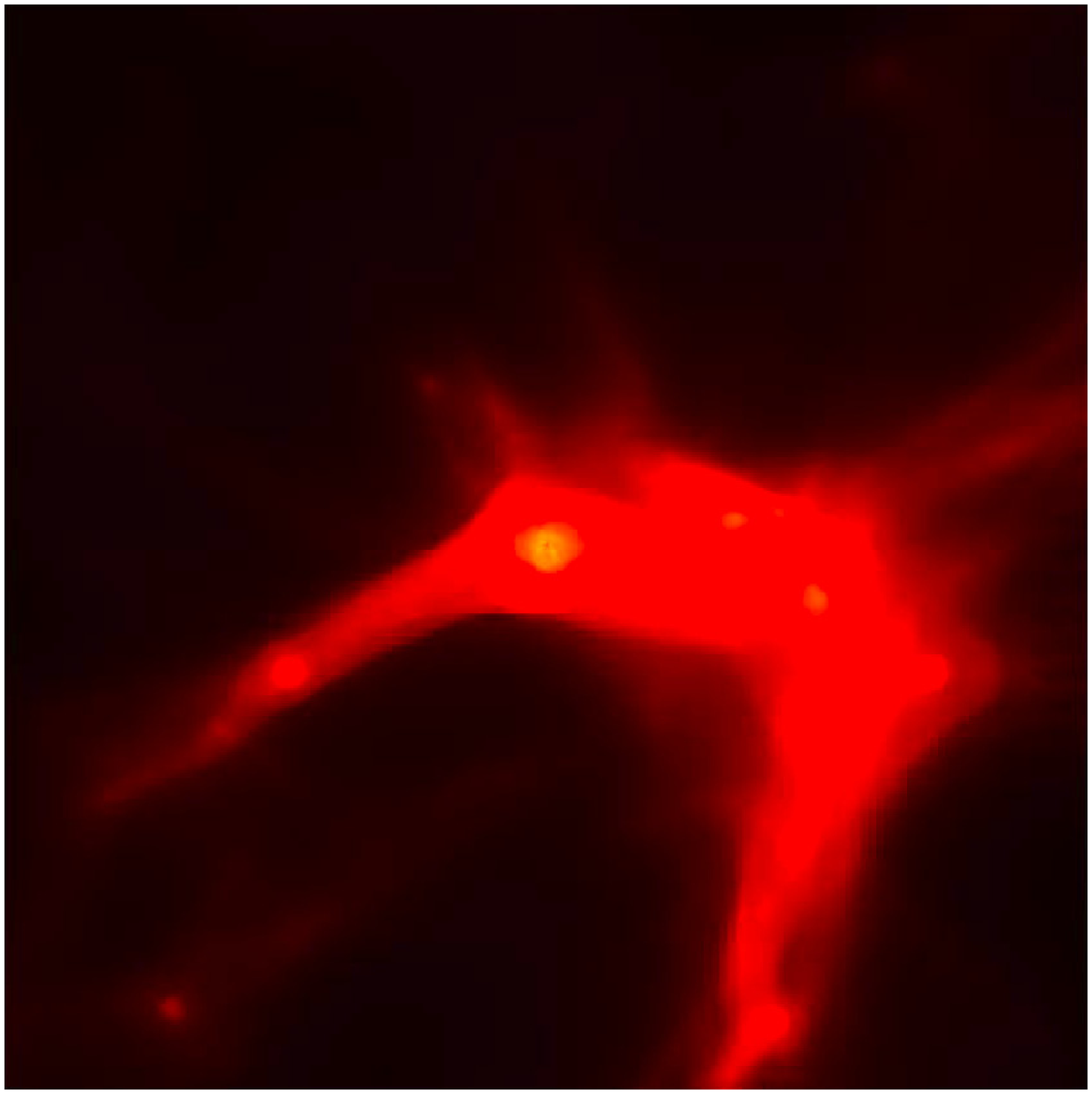,width=0.45\linewidth,clip=} & 
\epsfig{file=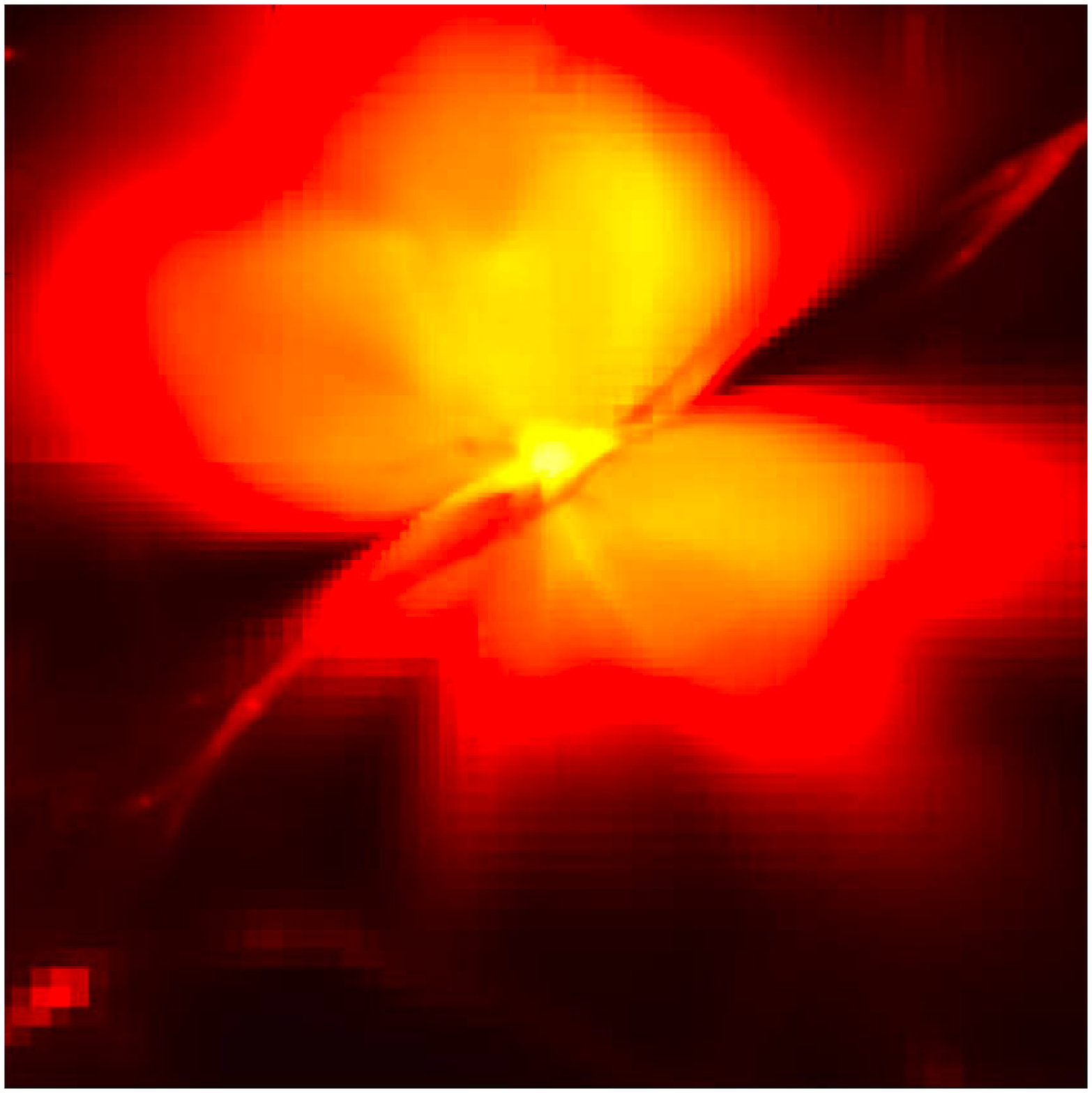,width=0.45\linewidth,clip=} \\
\end{tabular}
\end{center}
\caption{The environments of Pop III star formation and GRBs.  Left:  
temperature image of an 8.0 $\times$ 10$^5$ \Ms\ primordial halo at $z =$ 16.8.  
Right:  temperature image of the \HII\ region of a 100 \Ms\ Pop III star in the halo 
2 Myr later.  The image size is 4 kpc on a side.}
\label{fig:HII}
\vspace{0.1in}
\end{figure*}

The latest simulations suggest that Pop III GRBs may be produced by these
pathways more frequently than previously thought.  The discovery that 
fragmentation \citep[e.g.,][]{clark11,get12} and UV breakout \citep{hos11,
stacy12} in primordial halos may limit the masses of some Pop III stars to $
\lesssim$ 50 \Ms\ implies that more of them may fall into the mass range for 
GRBs than previously expected.  It is also now known that Pop III stars can 
die as compact blue giants that are susceptible to outbursts or as red 
supergiants that can enter a common envelope phase, depending on the 
degree of convective mixing \citep{wet12c} or rotational mixing \citep{yoon05,
yoon06} over the life of the star.  The fact that some Pop III stars are now 
known to form in binaries \citep{turk09} also improves the chances that some 
may enter a common envelope phase, a crucial ingredient for most collapsar 
scenarios. Finally, if many Pop III stars form with rotation rates that approach 
the breakup limit, as \citet{stacy11b} suggest, more GRBs may have occurred 
relative to the number of massive stars at very high redshifts than today 
\citep[although studies have shown that even at the critical velocity the cores 
of massive stars must often be spun up to even higher rates by a common 
envelope phase to produce collapsars;][]{fh05}.

\section{The Environments of Pop III GRBs}

\begin{figure}
\plotone{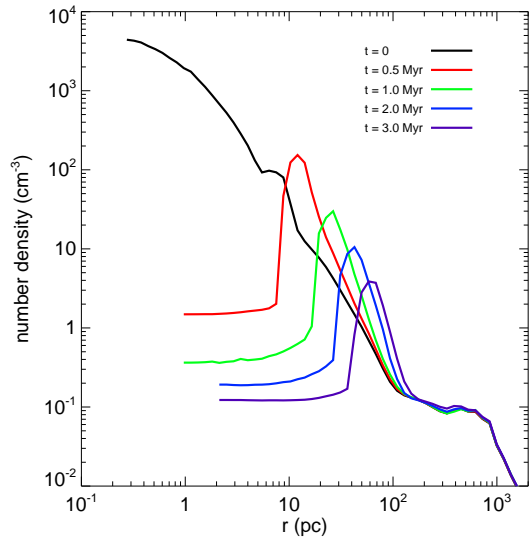} 
\caption{Spherically-averaged baryon density profiles for the \HII\ region of 
a 100 \Ms\ Pop III star.  Shocked, ionized core flows pile most of the baryons 
in the halo into a dense shell of radius $\sim$ 100 pc by the end of the life of 
the star.  Note that a second, smaller halo that is about to merge with the halo 
hosting the star is visible as the density bump at $\sim$ 10 pc at $t =$ 0, the 
time that the star is switched on.}
\label{fig:HIIprof}
\end{figure}

Because Pop III stars are very massive, they usually ionize the halos that gave 
birth to them, creating \HII\ regions 2.5 - 5 kpc in radius and driving all the gas
from the halo in shocked flows on timescales of $\sim$ 2 Myr \citep[e.g.,][]{
wan04}.  In Figure~1 we show the \HII\ region of a 100 \Ms\ star in a 8.0 $\times$ 
10$^5$ \Ms\ halo at $z =$ 16.8 that was simulated with the {\it Enzo} code. These 
flows create uniform density profiles with $n \sim$ 0.1 - 1 cm$^{-1}$ that extend 
50 - 100 pc from the star, as shown in Figure~\ref{fig:HIIprof}.  For this reason, 
and because Pop III stars are not thought to drive strong winds because they lack 
metals, previous studies have taken the GRB jet to propagate into uniform \HII\ 
region densities.  In the past, with less detailed observations, such profiles have 
yielded afterglow light curves that are in reasonable agreement with those of 
GRBs in the local universe. 

In reality, the ejection of the envelope and the fast winds that accompany most 
GRBs reset the environment in the vicinity of the star.  This holds true even for 
Pop III stars because the ejection of the envelope is driven by kinematics, not 
metallicity, and the He core would still drive a wind after the ejection because 
rotational and convective mixing enrich the core with metals.  If the progenitor 
is a single star with an LBV outburst, the envelope would be a fast wind driving 
a slow shell into a uniform \HII\ region.  Collimated flows can complicate this 
picture.  If less mass is blown along the axis of the star (and hence the jet) than 
its equator, the density along the axis will be intermediate to those of the \HII\ 
region and the shell.  Typical LBV outbursts expel 1 - 10 \Ms\ shells.

\begin{figure*}
\begin{center}
\begin{tabular}{cc}
\epsfig{file=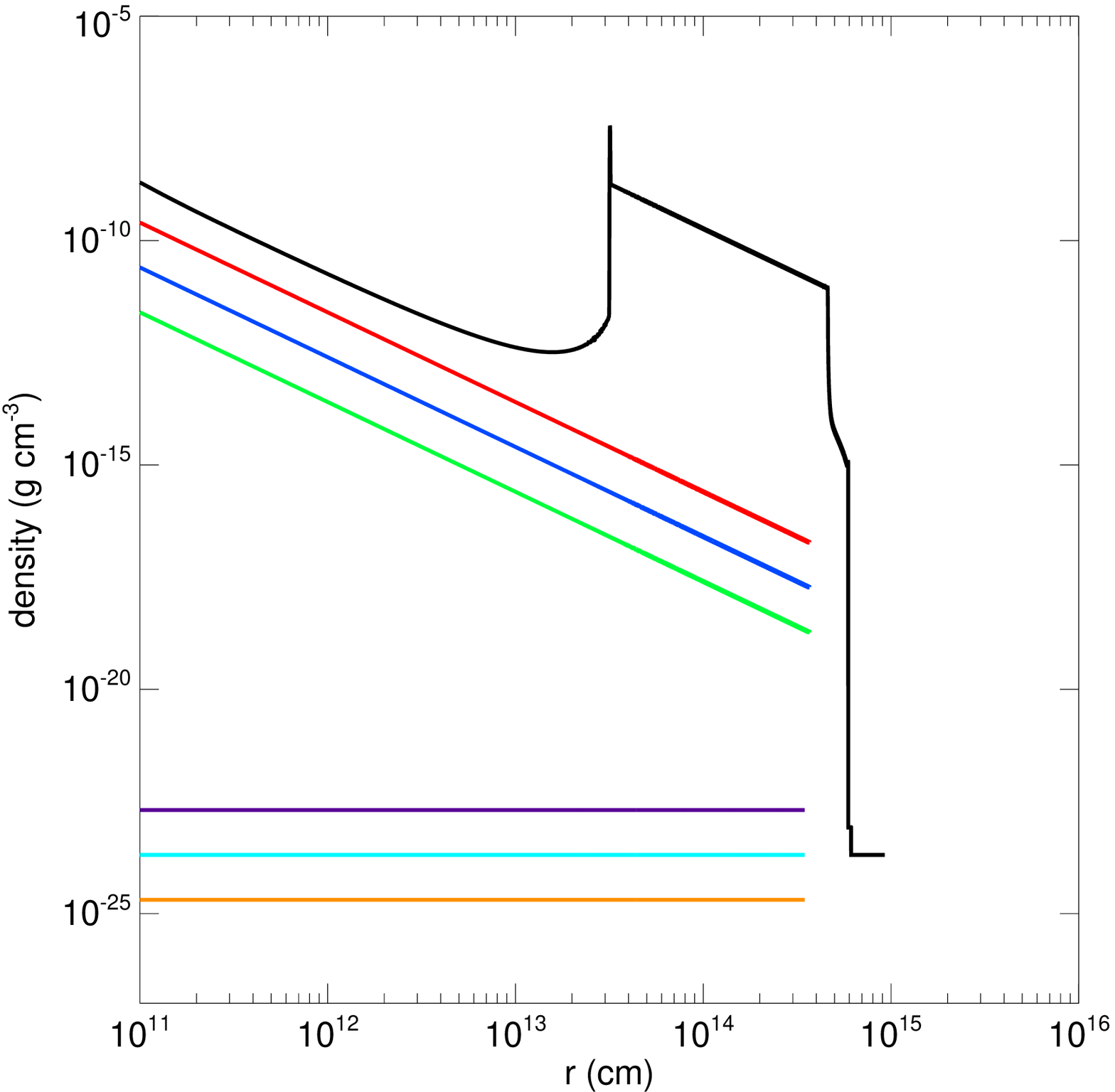,width=0.45\linewidth,clip=} & 
\epsfig{file=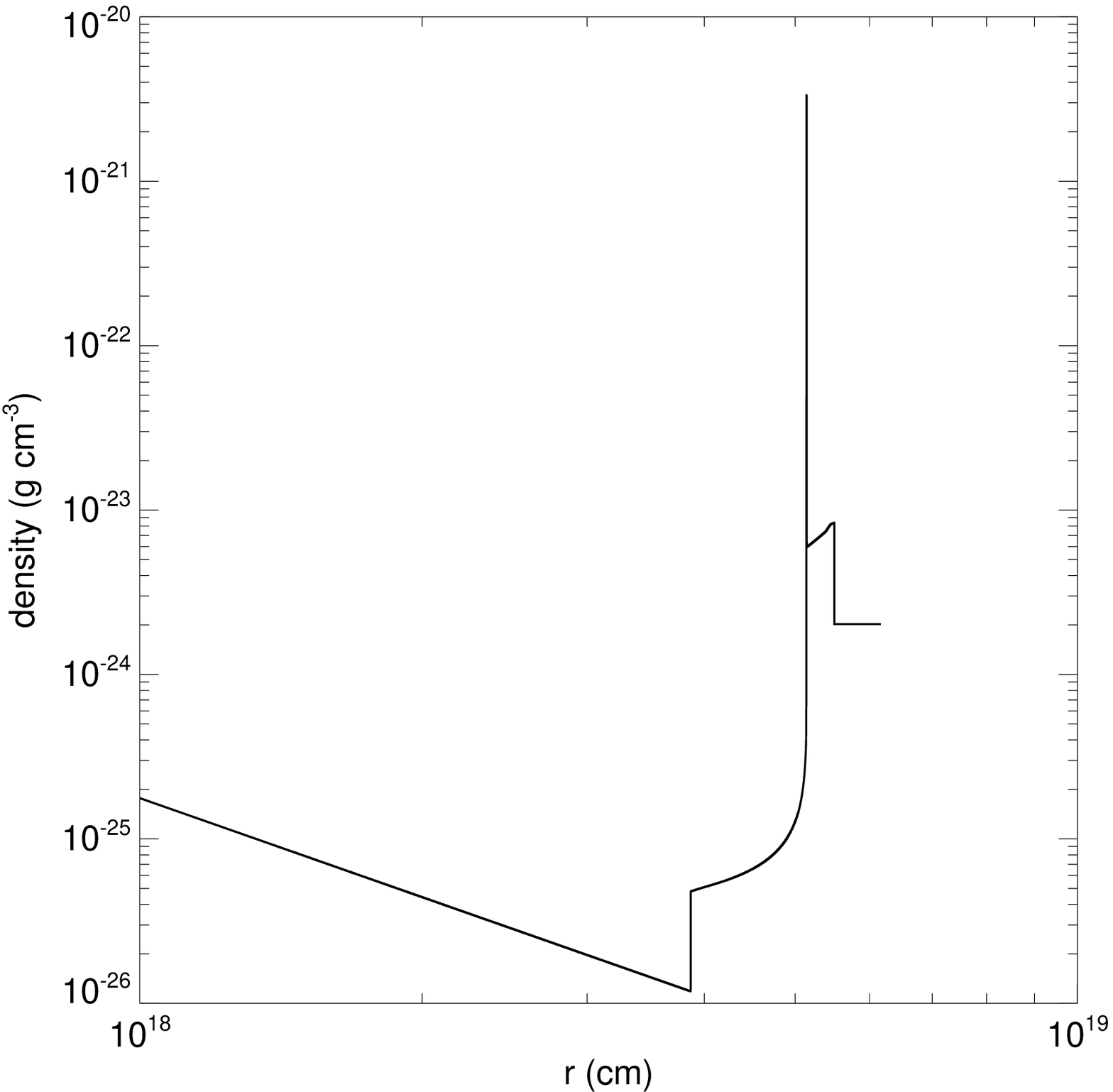,width=0.45\linewidth,clip=} \\
\end{tabular}
\end{center}
\caption{Circumburst profiles in our Pop III GRB survey.  Left:  density profiles for 
a 5 \Ms\ shell ejected by a He merger at 1.5 AU (black). Also included are 3 simple 
power-law winds (red: 10$^{-4}$ \Ms\ yr$^{-1}$; blue: 10$^{-5}$ \Ms\ yr$^{-1}$; 
green: 10$^{-6}$ \Ms\ yr$^{-1}$) and three uniform \HII\ regions (violet: 10 cm$^{
-3}$; cyan: 1.0 cm$^{-1}$; brown: 0.1 cm$^{-3}$). Right:  density profile of a 1 \Ms\ 
shell at $\sim$ 1 pc that was ejected in a binary merger 2000 yr prior to the burst.}
\label{fig:env}
\vspace{0.1in}
\end{figure*}

If the GRB is instead created by a merger in which the companion is a star, 
the envelope would just be a power-law wind profile, since in most cases the 
shell will be driven beyond the reach of the jet by the time of the burst.  If the 
companion is a BH or NS, the shell will be much closer to the star at the time 
of the burst.  Indeed, population synthesis models predict the average 
radius of shells expelled by He mergers to be 1 - 2 AU \citep[see Figure~11 
of][for $Z =$ 0.1 \Zs\ stars]{fb13}. However, the bulk of the envelope tends to 
be ejected along the equatorial plane of the progenitor in such events, so the 
density profile can again be intermediate to that of the \HII\ region and a fast 
wind pushing a massive shell \citep[see Figure~6 of][]{pas12}.  We note that 
the jet can encounter clumps even if little of the envelope is expelled along 
the axis of the burst because violent instabilities can form in the \HII\ region 
of the star that leave dense fragments in its vicinity at the time of its death 
\citep{wn08b,wn08a}.  

In sum, the environments of Pop III GRBs fall into four basic categories:

\begin{enumerate}
\item{the power-law density profile of a fast wind in which a shell has been 
driven beyond the reach of the GRB jet: \vspace{0.05in} 
\begin{equation}
\rho_\mathrm{w}(r) = \frac{\dot{m}}{4 \pi r^2 v_\mathrm{w}}, \vspace{0.05in}
\end{equation}
where $\dot{m}$ is the mass loss rate of the wind and $v_\mathrm{w}$ is its 
speed.  This profile is formed in most binary mergers between two stars and 
some single collapsars.}

\item{a fast wind driving a massive shell into a Pop III \HII\ region, where the 
shell is at $\sim$ 0.2 pc at the time of the burst.  This profile is created by  
some single collapsars and the relatively few mergers between two stars that 
place a shell within the range of the jet at the time of the burst.}

\item{a fast wind pushing into a massive shell in an \HII\ region in which the 
inner surface of the shell is at 1 - 2 AU at the time of the burst.  This is the
profile of a He merger.}

\item{a diffuse, uniform \HII\ region profile that a jet might encounter along 
certain lines of sight, such as in a toroidal mass ejection.  This envelope
is also appropriate for the much more massive and energetic GRBs 
considered by \citet{suwa11} and \citet{nsi12}, in which the star collapses
without ejecting a shell.}

\end{enumerate}

\section{Pop III GRB Models}

\begin{table*}[t]
\begin{center}
\begin{tabular}{|l|cc|c|c|}

\hline 
\multicolumn{1}{|c|}{Instrument}    & \multicolumn{2}{|c|}{Frequency Range}  &     Sensitivity    &  Integration Time  \\
                                    &     Min (GHz)     &     Max (GHz)      &        (mJy)       &      (minutes)     \\
\hline 

LWA                         &   $1.0\times10^{-2}$   &   $8.8\times10^{-2}$   &   $1.0\times10^{-1}$   &   $4.8\times10^{2}$  \\
LOFAR		        &   $3.0\times10^{-2}$   &   $8.0\times10^{-2}$   &   $8.0\times10^{-1}$   &   $4.8\times10^{2}$  \\
VLA                          &             $1.0$              &   $4.0\times10^{1}$    &   $2.5\times10^{-2}$   &   $3.6\times10^{2}$  \\
SKA                          &             $0.1$              &   $1.0\times10^{2}$    &   $5.0\times10^{-4}$   &   $1.7\times10^{1}$  \\
ALMA                       &   $8.4\times10^{1}$    &   $7.2\times10^{2}$    &   $3.5\times10^{-2}$   &   $3.6\times10^{2}$  \\
GRIPS IRT               &   $1.0\times10^{4}$    &   $4.0\times10^{4}$    &   $9.3\times10^{-4}$   &   $8.3\times10^{0}$  \\
{\it JWST} NIRcam   &   $6.3\times10^{4}$    &   $4.3\times10^{5}$    &   $1.0\times10^{-4}$   &   $1.7\times10^{2}$  \\
WFIRST                   &   $1.5\times10^{5}$    &   $4.0\times10^{5}$    &   $2.6\times10^{-4}$   &   $1.7\times10^{1}$  \\
JANUS NIRT            &   $1.8\times10^{5}$    &   $4.3\times10^{5}$    &   $6.7\times10^{-2}$   &   $8.0\times10^{0}$  \\
EXIST IRT               &   $1.3\times10^{5}$    &   $1.0\times10^{6}$    &   $2.3\times10^{-3}$   &   $5.0\times10^{-1}$ \\
SVOM MXT             &   $7.3\times10^{7}$    &   $1.5\times10^{9}$    &   $4.3\times10^{-3}$   &   $1.7\times10^{-1}$ \\
SVOM ECLAIRS      &   $9.7\times10^{8}$    &   $6.0\times10^{10}$  &   $3.9\times10^{-2}$   &   $1.7\times10^{1}$  \\
LOBSTER WF XRT  &   $1.2\times10^{8}$    &   $3.6\times10^{9}$    &   $5.6\times10^{-3}$   &  $5.0\times10^{-1}$ \\
JANUS XRFM          &   $1.2\times10^{8}$    &   $6.0\times10^{9}$    &   $7.0\times10^{-2}$   &   $5.0\times10^{-1}$ \\
EXIST HET              &   $1.2\times10^{9}$    &   $3.6\times10^{10}$  &   $1.9\times10^{-3}$   &   $1.7\times10^{0}$  \\
Swift BAT                 &   $3.6\times10^{9}$    &   $3.6\times10^{10}$  &   $1.9\times10^{-4}$   &   $1.7\times10^{0}$  \\
Fermi GBM              &   $2.4\times10^{9}$    &   $6.0\times10^{12}$  &   $3.1\times10^{-5}$   &   $1.7\times10^{0}$  \\

\hline 

\end{tabular}
\end{center}
\caption{Frequency ranges and sensitivities for some current and proposed radio, 
infrared, and X-ray instruments.  All sensitivities are at the $5\sigma$ level for the 
given integration time.   \label{tab:instr}}
\end{table*}   

We consider GRBs in 12 density profiles.  Three are simple winds, with $\dot{m} =$ 
10$^{-4}$, 10$^{-5}$ and 10$^{-6}$ \Ms\ yr$^{-1}$.  Three are profiles in which a 5 
\Ms\ shell is driven by a 10$^{-5}$ \Ms\ yr$^{-1}$ wind out to $\sim$ 0.2 pc in \HII\ 
region densities of 0.1, 1.0 and 10 cm$^{-3}$. Three are profiles for He mergers, in
which a 5 \Ms\ shell is driven by a 10$^{-5}$ \Ms\ yr$^{-1}$ wind into the same 3 
\HII\ regions above but only to a distance of $\sim$ 1.5 AU.  The last three profiles
are just the 3 \HII\ regions themselves.  In all cases, we take $v_\mathrm{w} =$ 
2000 km s$^{-1}$, $v_{\mathrm{shell}}$ = 200 km s$^{-1}$, and the composition of 
the envelopes to be primordial, 76\% H and 24\% He by mass. We show density 
profiles for all our models in Figure~\ref{fig:env}.

Much higher energies ($E_{\mathrm{iso},\gamma} =$ 10$^{55}$ - 10$^{57}$ erg) 
are sometimes invoked for Pop III GRBs in part because the mass of the star, and 
hence the reservoir of gas that is available to the central engine, is thought to be 
much greater than in stars today. \citet{suwa11} and \citet{nsi12} also find that 
such energies are required for the jet to punch through the outer layers of very 
massive stars that do not shed their hydrogen envelopes.  We proceed under the 
assumption that most Pop III GRBs are similar to those today and consider the 
usual energies for such events, 10$^{51}$, 10$^{52}$, and 10$^{53}$ erg.  There 
are thus a total of 36 models in our simulation campaign.  For simplicity, we take
the initial Lorentz factor of the jet, $\Gamma_0$, to be 500, and the duration of the 
burst in the Earth frame to be 100 seconds.  We take all the explosions in our study
to occur at $z =$ 20.

\subsection{ZEUS-MP Outburst Models}

We calculate density profiles for outbursts in Pop III \HII\ regions with ZEUS-MP
\citep{wn06} in the same manner as in M12.  We treat stellar winds and outbursts 
as time-dependent inflows at the inner boundary of a one-dimensional (1D) 
spherical grid:  
\vspace{0.05in}
\begin{equation}
\rho \, = \, \frac{\dot{m}}{4 \pi {r_{\mathrm{ib}}}^2 v_{\mathrm{w}}}, \vspace{0.05in}
\end{equation}
where $r_{\mathrm{ib}}$ is the radius at the inner boundary and $v_{\mathrm{w}}$ 
is the wind velocity.  Outbursts are generated by increasing $\dot{m}$ and lowering 
$v_{\mathrm{w}}$.  At the beginning of a run the grid is initialized with one of the 3
\HII\ region densities.  The mesh has 32,000 uniform zones and extends from 3.084 
$\times$ 10$^{10}$ cm to 9.252 $\times$ 10$^{14}$ cm ($\sim$ 60 AU) for He 
merger simulations and from 10$^{-5}$ pc to 0.2 pc for other mergers.  We impose 
outflow conditions on the outer boundary, and the grid is domain decomposed into 
8 tiles, with 4000 zones per tile and one tile per processor.

Radiative cooling can flatten shells into cold, dense structures that strongly affect 
the evolution of the GRB jet.  Although there are no metals or dust in our primordial 
ejections the shell can still cool by x-ray emission and H and He lines in the shocked 
gas.  Our ZEUS-MP models include collisional excitation and ionization cooling by H 
and He, recombinational cooling, H$_2$ cooling, and bremsstrahlung cooling, with 
our nonequilibrium H and He reaction network providing the species mass fractions 
(H, H$^{+}$, He, He$^{+}$, He$^{2+}$, H$^{-}$, H$^{+}_{2}$, H$_{2}$, and e$^{-}$) 
needed to calculate these collisional cooling processes.  The hydrodynamics in our
models is always evolved on the lesser of the cooling and Courant times to capture
the effect of cooling on the structure of the flow.  We neglect the effect of ionizing 
radiation from the progenitor on winds and shells.  This treatment is approximate, 
given that the star illuminates the flow and that its luminosity evolves over time.  
However, the energy deposited in the flow by ionizations is small in comparison to 
its bulk kinetic energy and is unlikely to alter its properties in the vicinity of the burst. 

One important difference between our shell profiles and those of M12 is that a fast 
wind does not precede the ejection of the envelope and later detach from its outer 
surface to create a very low-density rarefaction zone ahead of the shell.  The 
outburst instead plows up the much higher density \HII\ region, as shown in 
Figure~\ref{fig:env}.  Note that there are significant differences in the structures of 
shells ejected just before the burst and ejections 2000 yr before the burst. Radiative 
cooling has inverted the density profile inside the shell at later times and created a 
much larger density jump at the interface between the shell and the termination 
shock due to the wind that piles up at its inner surface. These structural differences, 
together with the distance of the shell from the explosion, have important 
consequences for GRB light curves as we discuss below.

\section{Pop III GRB Light Curves}

We tabulate frequency bands and sensitivities for current and proposed radio, NIR, and X-ray instruments in Table~\ref{tab:instr}.  
The radio instruments include the Long Wavelength Array \citep[LWA;][]{taylorEA12}, the Low Frequency Array 
(LOFAR)\footnote{http://www.astron.nl/radio-observatory/astronomers/lofar-imaging-capabilities-sensitivity/sensitivity-lofar-array/sensiti}, 
the Very Large Array (VLA)\footnote{https://science.nrao.edu/facilities/vla/docs/manuals/propvla/
determining/source}, the Square Kilometer Array 
(SKA)\footnote{http://www.astron.nl/radio-observatory/astronomers/lofar-imaging-capabilities-sensitivity/sensitivity-lofar-array/sensiti}, and 
the Atacama Large Millimeter Array (ALMA)\footnote{http://almascience.eso.org/proposing/sensitivity-calculator}. The NIR instruments are 
the GRB Investigation via Polarimetry and Spectroscopy Infrared Telescope \citep[GRIPS IRT;][]{greinerEA12}, the {\it James Webb Space 
Telescope} ({\it JWST}) NIR Camera (NIRCam)\footnote{http://www.stsci.edu/jwst/instruments/nircam/sensitivity}, the Wide-Field Infrared 
Survey Telescope \citep[WFIRST;][]{spergelEA13}, the Joint Astrophysics Nascent Universe Satellite NIR Telescope \citep[JANUS NIRT;][]{
Burrows10}, and the Energetic X-ray Imaging Survey Telescope Infrared Telescope \citep[EXIST IRT;][]{grindlay10}. The X-ray instruments 
include the Space-based Variable Objects Monitor (SVOM) Micro-channel X-ray Telescope (MXT) and ECLAIRs \citep{paulEA11}, the Large
Angle Observatory with Energy Resolution (LOBSTER) Wide-Field X-ray Telescope \citep[WF XRT;][]{goren11}, the JANUS X-ray Flash 
Monitor \citep[XRFM;][]{falconeEA09}, the EXIST High-Energy Telescope \citep[HET;][]{hongEA09}, and the {\it Swift} Burst Alert Telescope 
(BAT) and {\it Fermi} Gamma-ray Burst Monitor (GBM)\footnote{http://fermi.gsfc.nasa.gov/science/mtgs/symposia/2012/
program/tue/PJenke.pdf}.

\subsection{\HII\ Regions}

\begin{figure}
\begin{center}
\begin{tabular}{c}
\epsfig{file=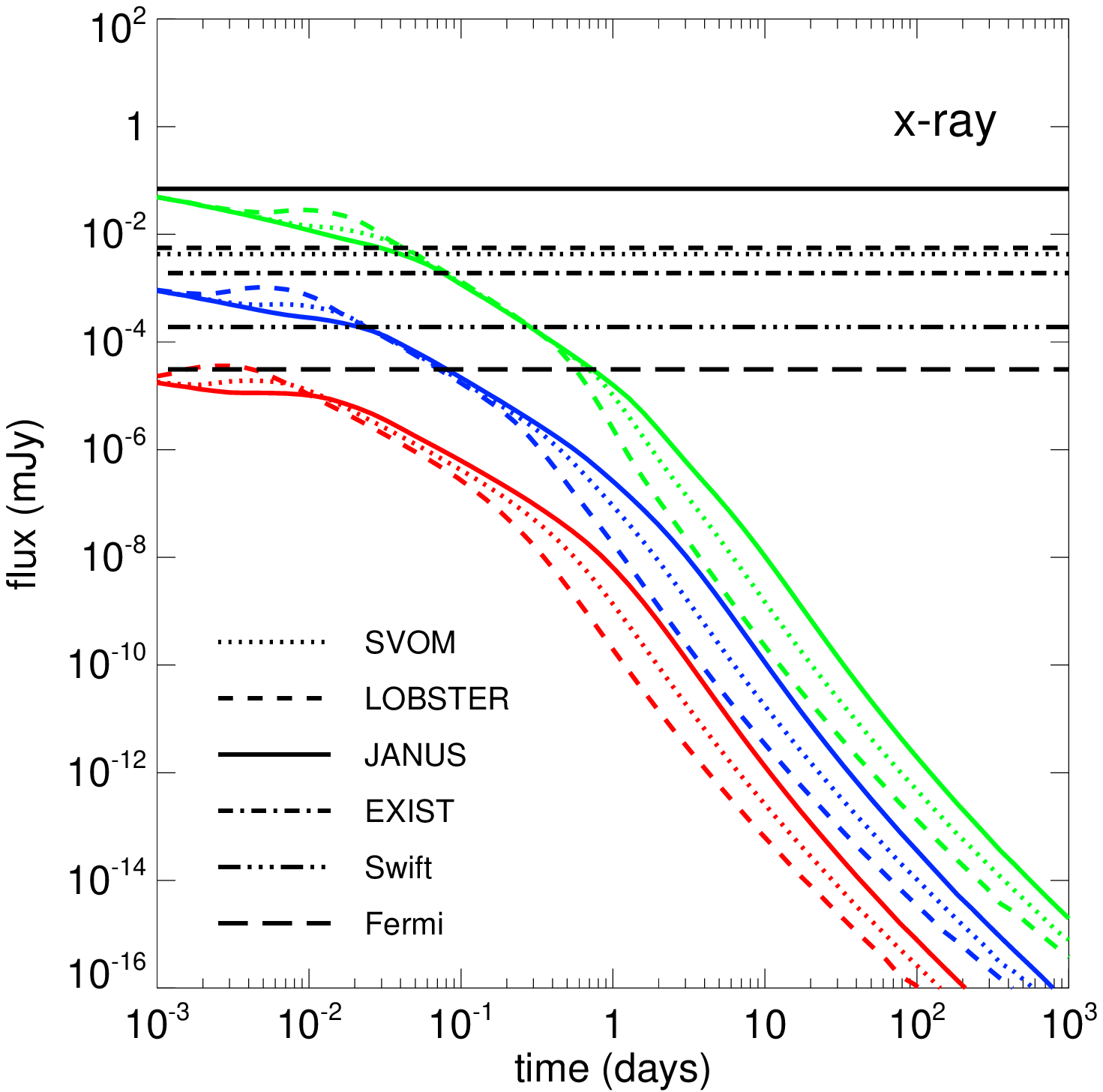,width=0.85\linewidth,clip=} \\ 
\epsfig{file=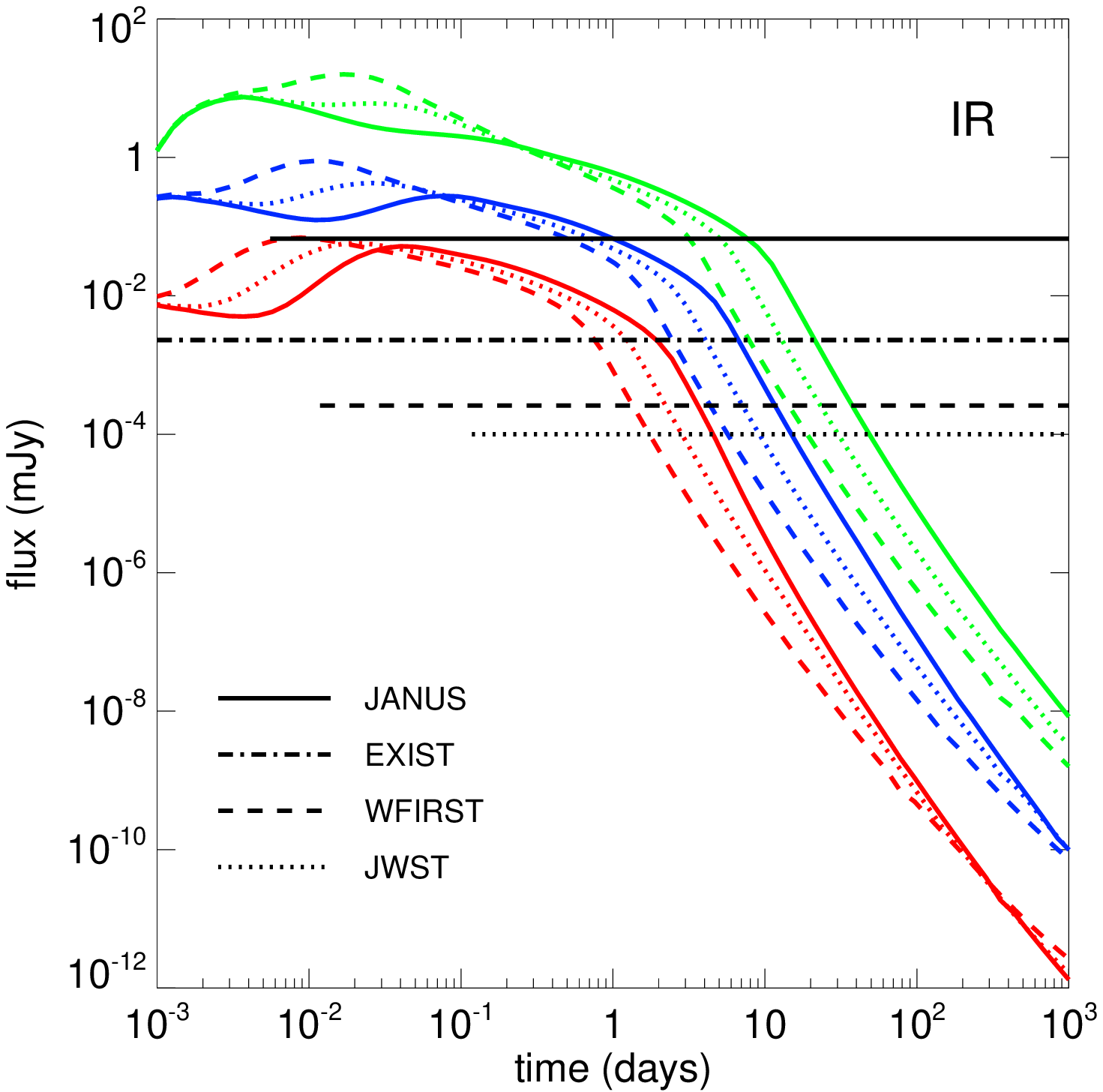,width=0.85\linewidth,clip=} \\
\epsfig{file=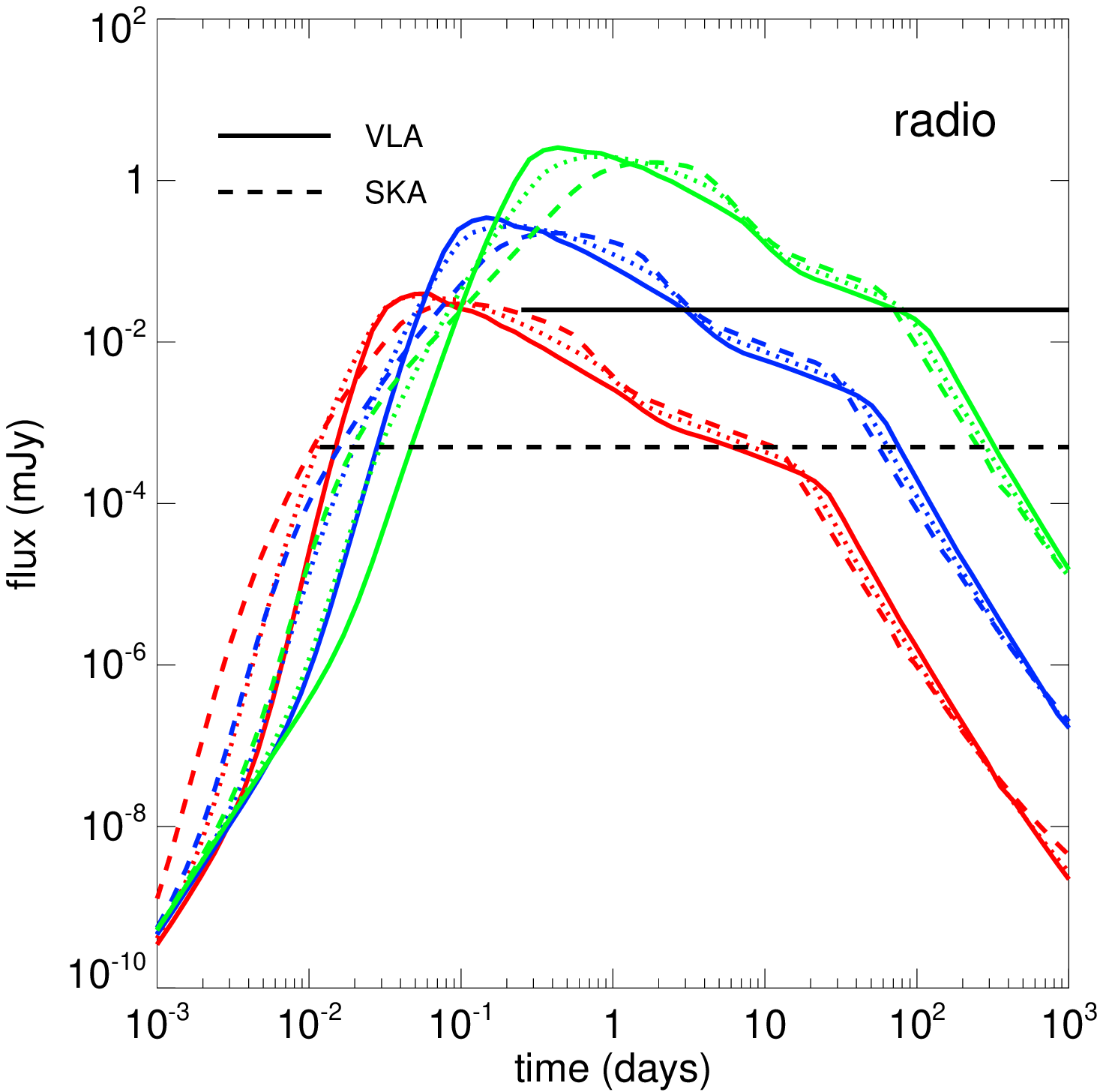,width=0.85\linewidth,clip=} 
\end{tabular}
\end{center}
\caption{Afterglow light curves for Pop III GRBs in uniform \HII\ regions for $
E_{\mathrm{iso},\gamma} =$ 10$^{51}$ erg (red), 10$^{52}$ erg (blue), and 
10$^{53}$ erg (green).  Top: X-ray (82.7 keV); center: NIR (3.0 $\mu$m); 
bottom: radio (5 GHz).  Solid:  $n =$ 0.1 cm$^{-1}$;  dotted:  $n =$ 1.0 cm$
^{-1}$; dashed:  $n =$ 10 cm$^{-1}$.  All times are in the earth frame.}
\label{fig:HII}
\end{figure}

We show afterglow light curves for Pop III GRBs in uniform-density \HII\ regions 
at $z =$ 20 in Figure~\ref{fig:HII}.  In all three plots sensitivity limits are shown 
for the appropriate instruments, beginning at the minimum integration time for 
each one that would result in a detection.  The peak flux occurs at later times for 
lower frequencies.  Fluxes are highest in the NIR, reaching $\sim$ 10 mJy for a 
$10^{53}$ erg burst in an $n = 10$ cm$^{-3}$ \HII\ region.  They are somewhat 
lower in the radio and X-ray, although the radio flux falls off only gradually after 
reaching its peak.  In the NIR and X-ray, the flux scales roughly with the burst 
energy, but in the radio the flux increases by a factor of $\sim$ 50 with each 
decade in energy.  Also, the NIR and X-ray afterglows are brightest in the 
highest densities at early times, but after a few hours this trend is reversed.  
There is much less variation in flux with ambient density in the radio afterglows.

The peak flux at a given frequency is greatest in the NIR at $\sim$ 3 $\mu$m, 
and it decreases monotonically above and below this frequency.  Pop III GRB 
afterglows in \HII\ regions will thus most easily be detected in current and 
proposed NIR instruments.  The afterglow reaches a peak at or before $\sim 
10^{-1}$ days but falls off gradually enough that instruments such as GRIPS 
and WFIRST could detect a $10^{51}$ erg burst for up to about two days and 
a $10^{53}$ erg burst for up to 10 days.  The JANUS NIRT would detect a $10
^{53}$ erg burst for about ten days, while a $10^{51}$ erg burst would be just 
at the threshold of detectability.  The EXIST IRT would see NIR afterglows for 
0.5 - 25 days at $z \sim$ 20, and {\it JWST} could detect these events for 2 - 
80 days, depending on energy and \HII\ region density.

In the radio portion of the spectrum, the VLA could just barely detect a $10^{51}
$ erg burst in a $0.1$ cm$^{-3}$ \HII\ region, while the the same burst would be 
visible to the SKA for about an hour.  At the opposite extreme, a $10^{53}$ erg 
GRB in a $10$ cm$^{-3}$ \HII\ region would be detectable by the VLA for nearly 
80 days while SKA would see the same afterglow for $\sim 200$ days.  Because 
the peak flux is reached earlier at higher frequencies, it will be difficult for ALMA 
to detect even the brightest afterglow for more than about a day. LOFAR will not
see these explosions because its sensitivity falls below their peak fluxes. At $80$ 
MHz, the peak flux only reaches $\sim10^{-2}$ mJy, well below the LOFAR $5
\sigma$ sensitivity of $80$ mJy for an eight hour integration.
 
GRBs in \HII\ regions do not produce very bright X-ray afterglows.  The X-ray 
instruments that are sensitive to the lowest frequency X-rays will be the most 
suitable for detecting these events.  The MXT, HET, and WFI aboard SVOM, 
EXIST, and LOBSTER, respectively, plus the GBM on Fermi, with their high 
sensitivities and low minimum observable frequencies, will most easily detect 
Pop III GRBs.  Only the bursts with the highest energies would be visible.  A 
$10^{51}$ erg GRB would reach a maximum flux of only $\sim10^{-4}$ mJy 
at $3 \times 10^9$ GHz, well below the detection thresholds of every current
and proposed X-ray mission except for the Fermi GBM.  But a $10^{53}$ erg 
event could produce an afterglow with a flux of $\sim10^{-1}$ mJy at $10^9$ 
GHz immediately after the end of the prompt emission phase.  This flux would 
then fall off slowly enough for SVOM, EXIST, and LOBSTER to detect the 
afterglow for nearly a day.  We note that the FERMI LAT is not well suited to 
hunting for high-redshift GRB afterglows because of its high threshold 
frequency, $4.8\times10^{12}$ GHz, and modest sensitivity. 

\subsection{Winds}

\begin{figure}
\begin{center}
\begin{tabular}{c}
\epsfig{file=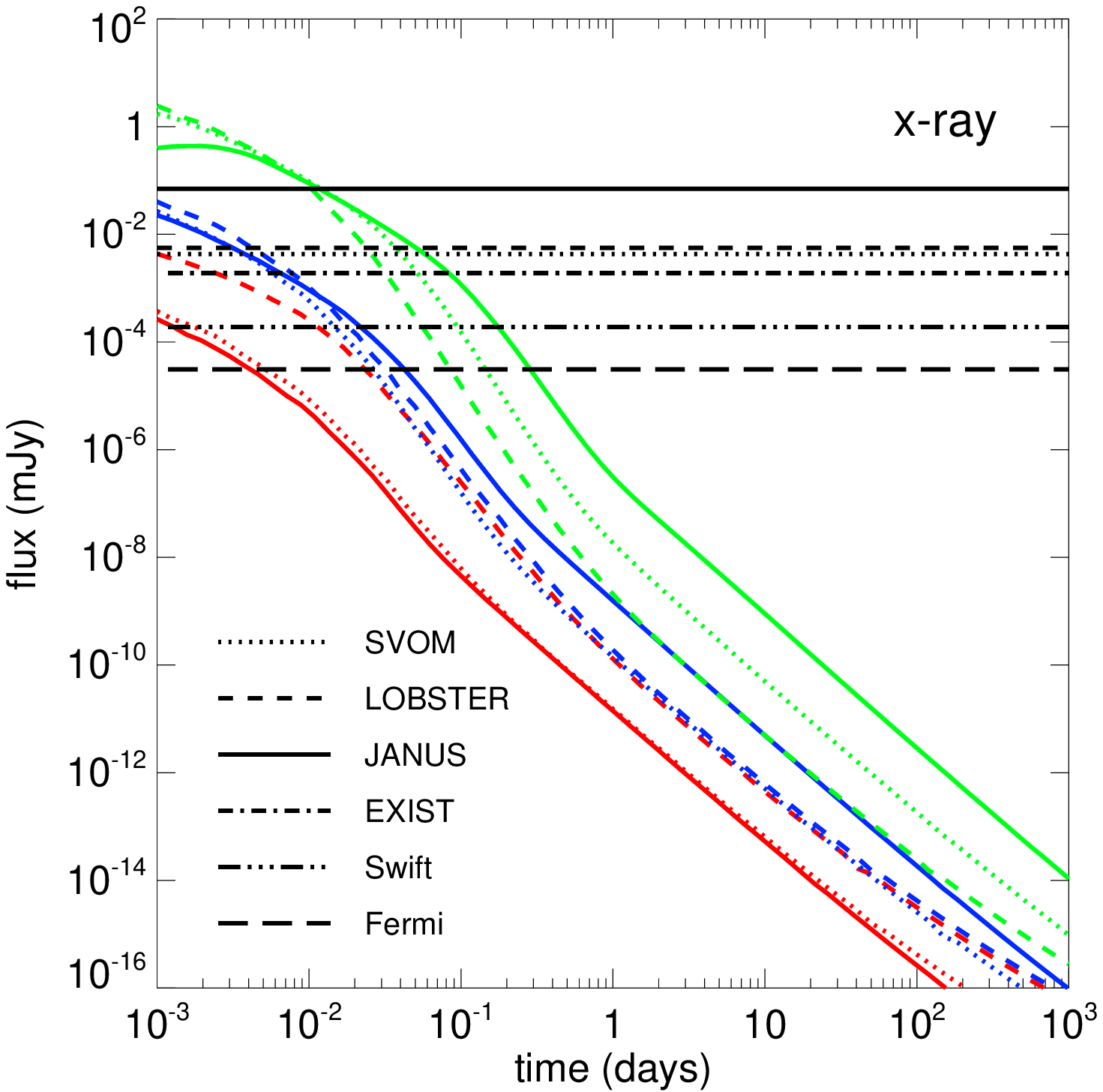,width=0.85\linewidth,clip=}  \\
\epsfig{file=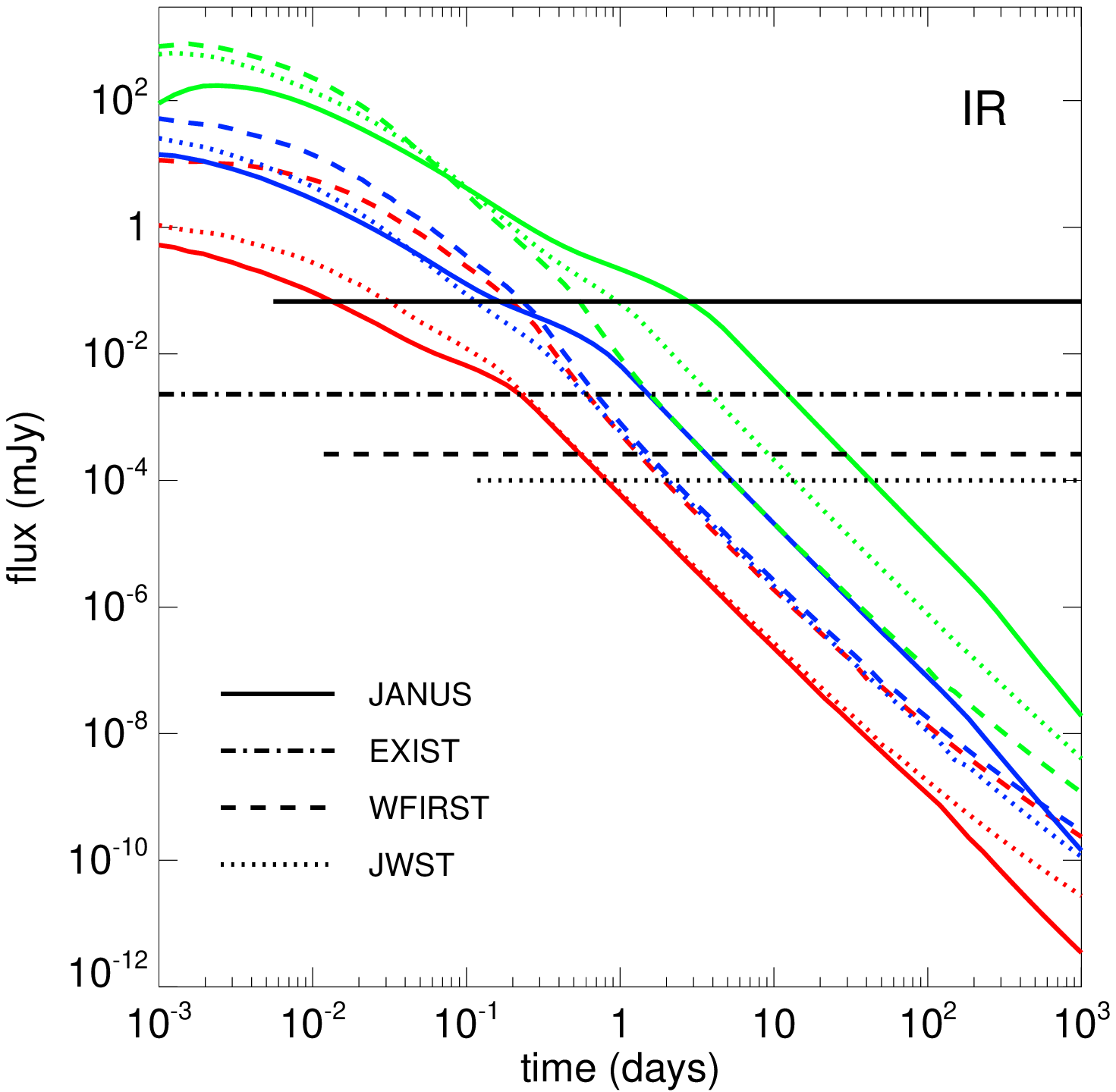,width=0.85\linewidth,clip=}  \\ 
\epsfig{file=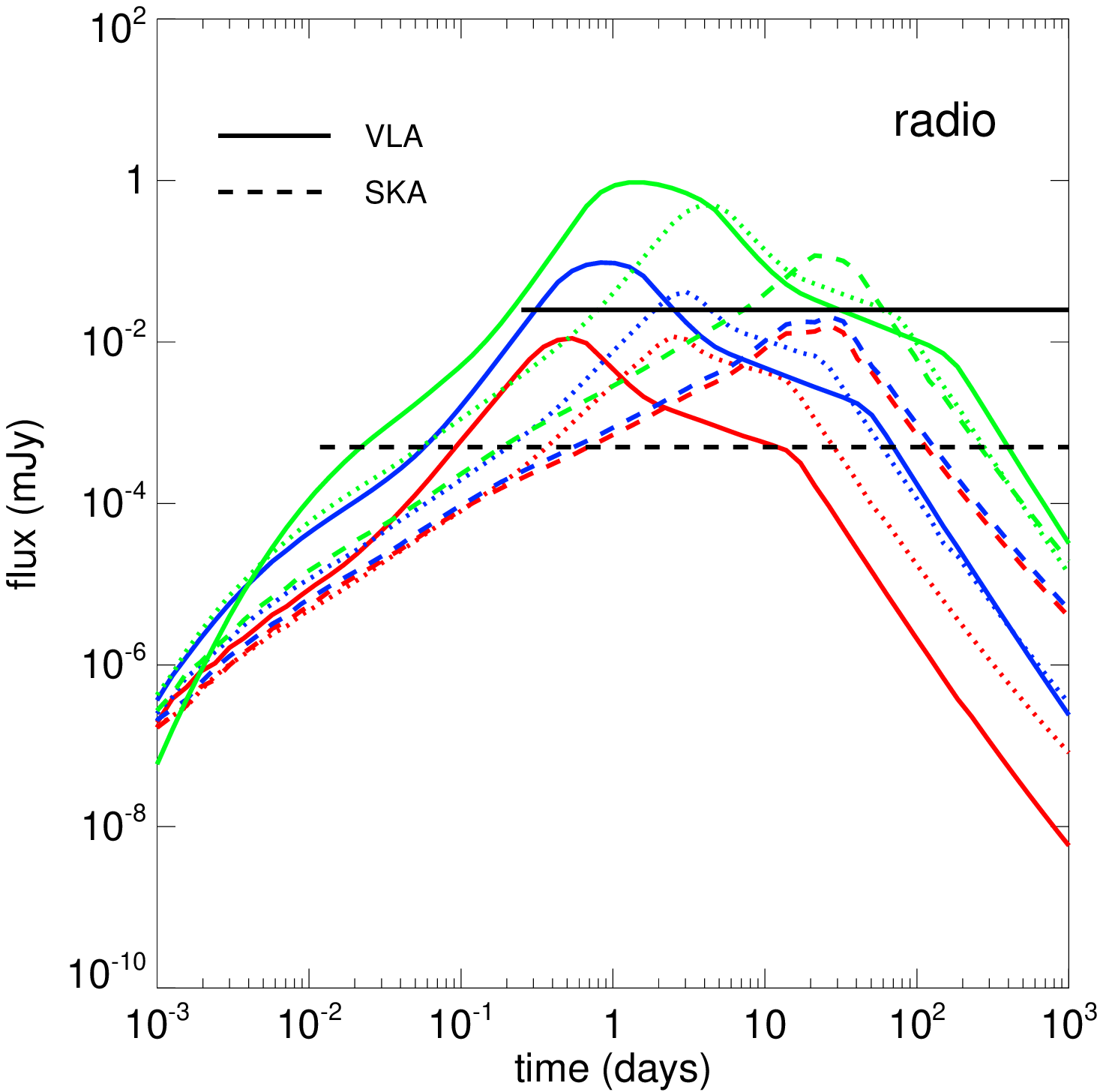,width=0.85\linewidth,clip=}  
\end{tabular}
\end{center}
\caption{Afterglow light curves for Pop III GRBs in $r^{-2}$ wind profiles for 
$E_{\mathrm{iso},\gamma} =$ 10$^{51}$ erg (red), 10$^{52}$ erg (blue), 
and 10$^{53}$ erg (green).  Top: X-ray (82.7 keV); center: NIR (3.0 $\mu
$m); bottom: radio (5 GHz).  Solid:  $\dot{m} =$ 10$^{-6}$ \Ms/yr; dotted: 
$\dot{m} =$ 10$^{-5}$ \Ms/yr; dashed:  $\dot{m} =$ 10$^{-4}$ \Ms/yr.  All 
times are in the earth frame.}
\label{fig:winds}
\end{figure}

Light curves for GRBs in $r^{-2}$ winds are shown in Figure \ref{fig:winds}.  
As noted earlier, these environments are expected in many binary merger 
scenarios in which the hydrogen envelope of the star has been driven by 
stellar winds to radii that are beyond the reach of the jet before its afterglow 
dims below visibility.  Like GRBs in \HII\ regions, afterglows from explosions in 
stellar winds are brightest in the NIR at about 3.0 $\mu$m, with peak fluxes 
falling off above and below this frequency.  However, due to the higher 
densities at smaller radii, afterglow fluxes at frequencies above the radio are 
greater in winds than in \HII\ regions for a given burst energy.  But fluxes in 
the radio are dimmer in winds than in \HII\ regions because they peak at later 
times, when the jet is at larger radii and lower densities than the \HII\ regions.  
All three fluxes scale roughly with the energy of the burst after a few hours, 
when the brightest afterglows are generally in the most diffuse envelopes.

A $10^{53}$ erg burst in a $10^{-4}$ \Ms\ yr$^{-1}$ wind reaches a maximum 
NIR flux of $\sim1$ Jy 100 seconds after the burst, making it readily detectable 
for up to 10 days with EXIST.  GRIPS and WFIRST could extend this window 
out to 10 - 20 days, and this event would be visible to JANUS NIRT for almost 
a day.  JANUS, however, with its somewhat lower sensitivities, would not be 
able to see the lowest energy bursts in the NIR because their fluxes fall so
quickly after the burst.  But EXIST would see these GRBs for up to half a day.  
The most sensitive of the instruments, the {\it JWST NIRCam}, could observe 
the least energetic GRBs for a day and the most energetic ones for 100 days, 
providing detailed followup to an initial detection in X-rays.

Besides being somewhat dimmer than in \HII\ regions, radio afterglows in 
winds reach peak fluxes at later times, 1 - 30 days instead of 0.1 - 3 days. 
The radio flux also peaks at later times in higher mass loss rates.  A $10^{
53}$ erg burst in a $10^{-6}$ \Ms\ yr$^{-1}$ stellar wind reaches a peak 
flux of $\sim$ 1 mJy at 5 GHz after 0.8 days.  It would be possible to detect 
such an afterglow with the VLA from about 0.1 - 80 days after the burst, 
and SKA would extend this range out to 200 days.  Neither ALMA nor 
LOFAR would be able to detect a Pop III GRB afterglow at $z = 20$ in a 
stellar wind.  The least energetic GRBs in the strongest winds are 
marginally detectable by VLA for a day and would be visible to SKA for 
about 10 days.

X-ray afterglows for GRBs in stellar winds are much brighter than those 
\HII\ regions.  A $10^{53}$ erg GRB at $z = 20$ produces fluxes greater 
than $20$ mJy at $10^9$ GHz for 2.5 hours.  This emission would be 
easily detectable by SVOM, EXIST, LOBSTER, JANUS, Swift, and the {\it 
Fermi} GBM.  The afterglow would be visible to the JANUS XRFM for 
nearly a half an hour and to {\it Fermi} for almost two hours.  At the other 
extreme, a $10^{51}$ erg GRB in a $10^{-6}$ \Ms\ yr$^{-1}$ wind would 
produce an afterglow that would only reach $\sim5\times10^{-3}$ mJy at 
$10^9$ GHz.  This event would only be detectable for $\sim200$ seconds 
with the \emph{Fermi} \emph{GBM}, and would likely not be found by 
other instruments.     

\subsection{He Mergers}

\begin{figure}
\begin{center}
\begin{tabular}{c}
\epsfig{file=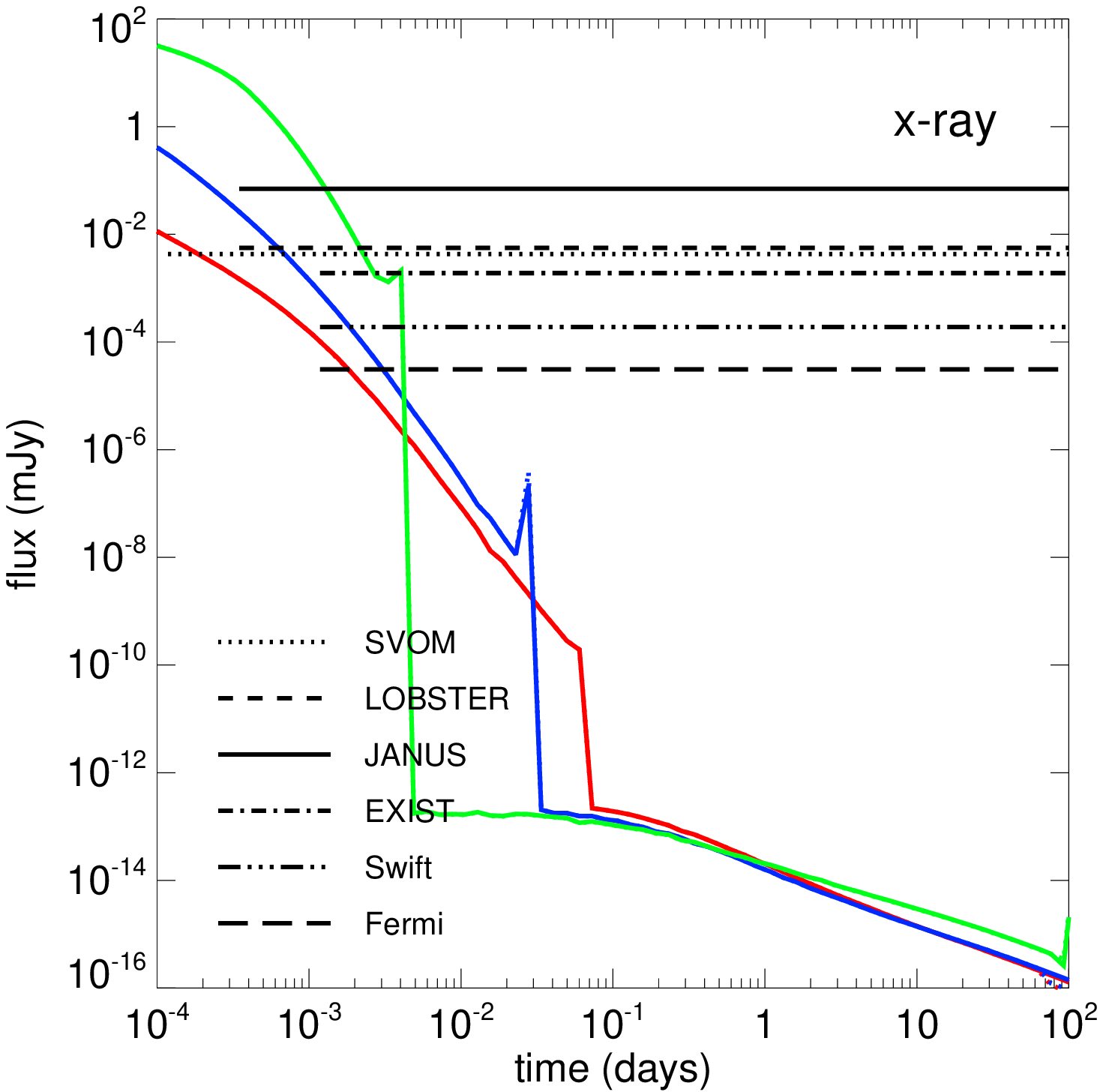,width=0.85\linewidth,clip=}  \\
\epsfig{file=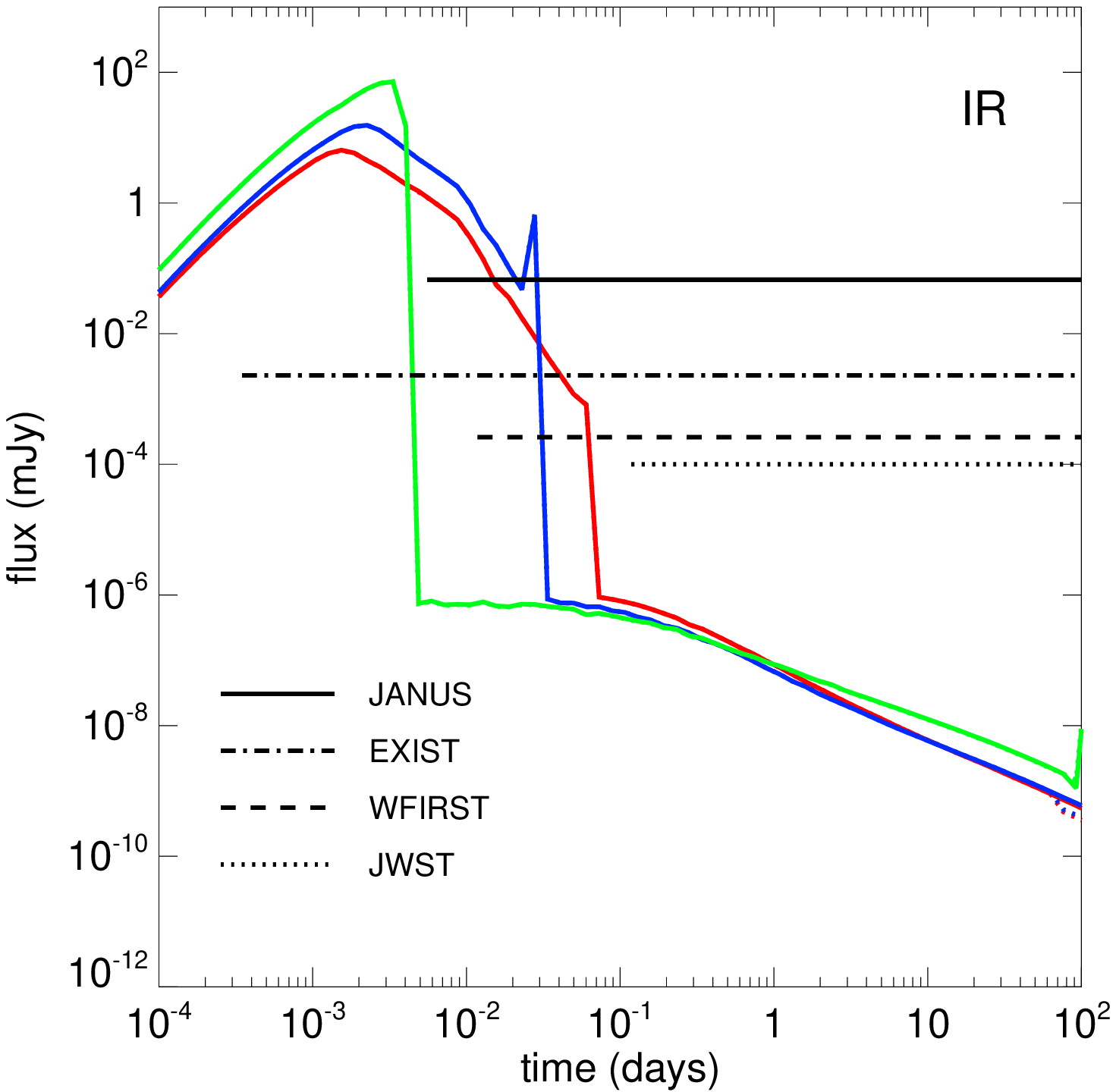,width=0.85\linewidth,clip=}  \\ 
\epsfig{file=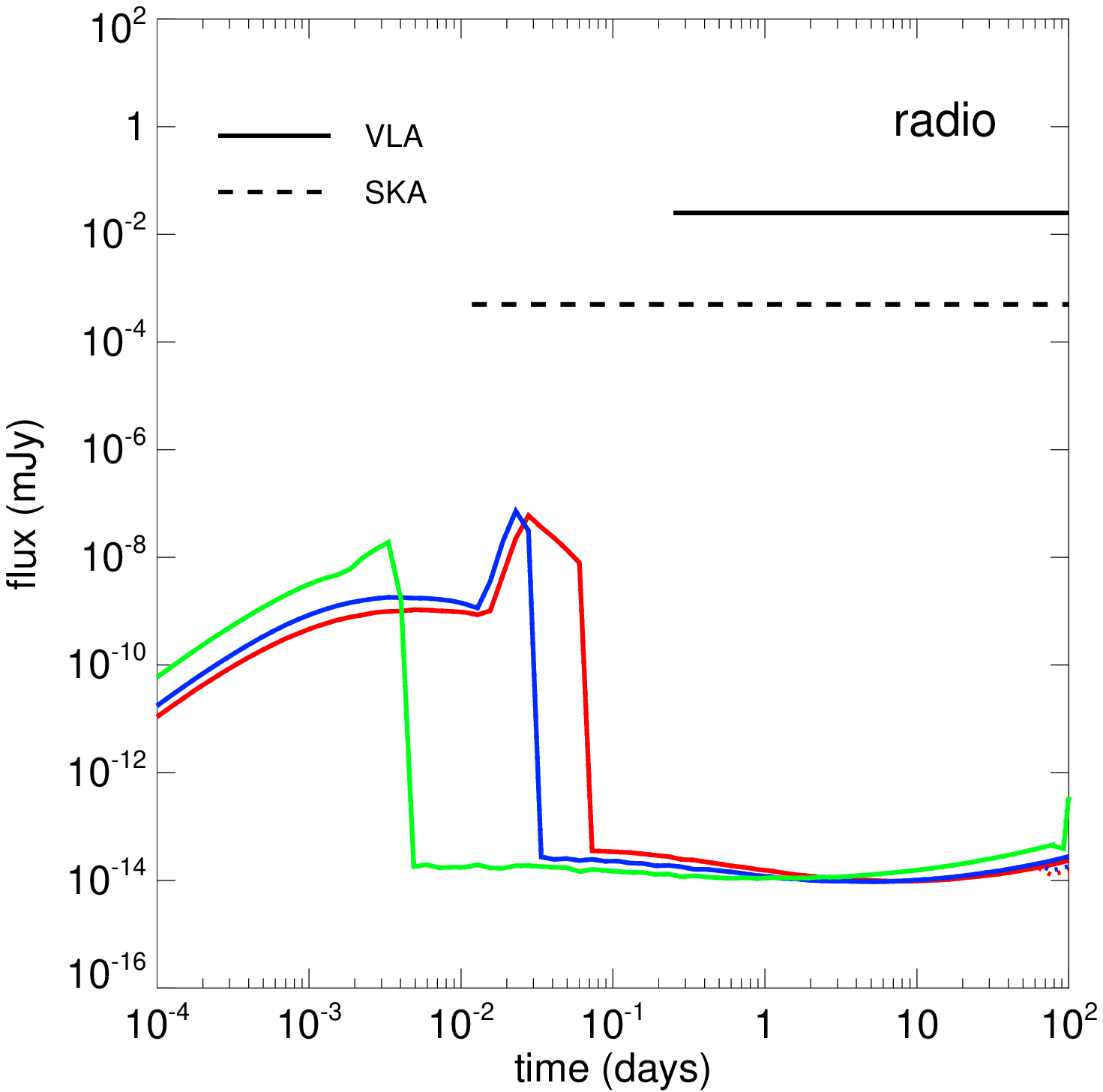,width=0.85\linewidth,clip=}  
\end{tabular}
\end{center}
\caption{Afterglow light curves for Pop III GRBs in 5 \Ms\ shells at 2 AU that 
were ejected in He mergers.  $E_{\mathrm{iso},\gamma} =$ 10$^{51}$ erg 
(red), 10$^{52}$ erg (blue), and 10$^{53}$ erg (green).  Top:  X-ray (82.7 
keV); center:  NIR (3.0 $\mu$m); bottom: radio (5 GHz).  Solid, dotted and 
dashed lines are for shells expanding in ambient densities $n =$ 0.1, 1.0, 
and 10 cm$^{-1}$, respectively.  All times are in the earth frame.}
\label{fig:HeM}
\end{figure}

We show X-ray, NIR and radio light curves for Pop III GRB jets crashing into 
massive shells ejected by helium merger events in Figure~\ref{fig:HeM}.  In
each case the 5 \Ms\ shell, which has been driven into \HII\ region densities
of 0.1, 1, or 10 cm$^{-3}$ by a stellar wind with $\dot{m} =$ 10$^{-3}$ \Ms\ 
yr$^{-1}$, has a radius of $\sim$ 2 AU.  The GRB jet breaks out into the $r^
{-2}$ wind before colliding with the shell.  The high densities in the wind at 
these small radii decelerate the jet to mildly-relativistic speeds in $\sim10$ 
minutes in the local frame.  Because the time between the ejection of the 
shell and the burst is short, the wind has not had time to pile up and form a 
termination shock at the inner layer of the shell. The jet therefore transitions 
directly from the wind to the hydrogen shell. 

A mildly relativistic reverse shock forms at the contact discontinuity between 
the wind and the shell, which then steps back through the jet in the frame of 
the jet.  A new forward shock develops with a lower Lorentz factor than that 
of the original forward shock but that is still, in general, mildly relativistic. The 
new forward shock advances into the shell and immediately encounters 
another density jump, where it produces another pair of forward and reverse 
shocks. This cycle continues, with several to dozens of shock pairs eventually 
being formed.  By the time the first forward shock reaches the interior of the 
shell, where the density is nearly constant and no more shocks are produced, 
all forward shocks have become non-relativistic and all reverse shocks have 
retreated through the forward shocks that created them.  

The jet remains in the shell for the entire 100 day simulation.  When it finally
emerges, there may be a slight re-collimation of the jet due to the low density 
of the surrounding medium.  Any rebrightening of the light curve would be 
minimal however, due to the low density of the medium.  We find that the \HII\
region density beyond the shell has essentially no effect on the afterglow light 
curve because the structure of the shell has not yet been altered by the relic 
\HII\ region. 

Light curves for Pop III GRBs in He merger shells are quite different from those 
in winds and \HII\ regions.  Their structure in a given band varies strongly with 
the energy of the burst and also across the bands themselves. They also 
exhibit much more variation over time, with sharp drops that are sometimes 
preceded by flares.  The peak flux is again in the NIR, but there is also a large 
initial X-ray flux.  Radio emission is suppressed by synchrotron self-absorption 
in the high densities in the shell, so detecting He merger GRBs at any frequency 
below the IR will be nearly impossible with current or proposed instruments. The 
flux at all frequencies is essentially quenched a short time after the jet enters the 
shell.  

The high densities in the shell (black plot in the left panel of Figure~\ref{fig:env}) 
produce a large NIR flux. A $10^{51}$ erg GRB reaches a peak flux of $10$ mJy 
at 150 seconds at 3.0 $\mu$m, while a $10^{53}$ erg GRB peaks at nearly $100$ 
mJy. Although they are visible to all the NIR detectors in Table \ref{tab:instr}, they 
will probably only be detected by satellites whose onboard X-ray instruments are 
triggered by the event because their NIR fluxes are so short-lived.  There would 
not be enough time to slew any ground-based instruments to capture the NIR 
afterglows of GRBs due to He mergers. This problem would be mitigated in cases 
in which there is more time between the ejection and the burst, since delays of up
to several days would not alter the structure of the shell but would allow followup
in the NIR from the ground.  Note that fluxes from more energetic bursts are 
quenched sooner because the jet reaches the shell in less time.  Higher burst 
energies imply larger ejecta masses for a given Lorentz factor, and they are not
decelerated by the wind as much as jets with lower masses.  

At 83 keV, the flux from a $10^{53}$ erg burst reaches $10$ mJy just after the 
end of the prompt emission phase.  This GRB would be visible to SVOM, JANUS,
LOBSTER, EXIST, and {\it Fermi} until the jet collides with the shell at $t_\text{obs} 
\lesssim 0.1$ days.  The flux then drops by many orders of magnitude, in some 
cases after producing a flare that lasts for a few seconds to a few minutes.  GRB 
afterglows with strong, transient X-ray and IR fluxes that end within a few hours 
would be clear signatures of a Pop III GRB from a He merger, particularly if there 
are flares.

\subsection{Binary Mergers}

\begin{figure}
\begin{center}
\begin{tabular}{c}
\epsfig{file=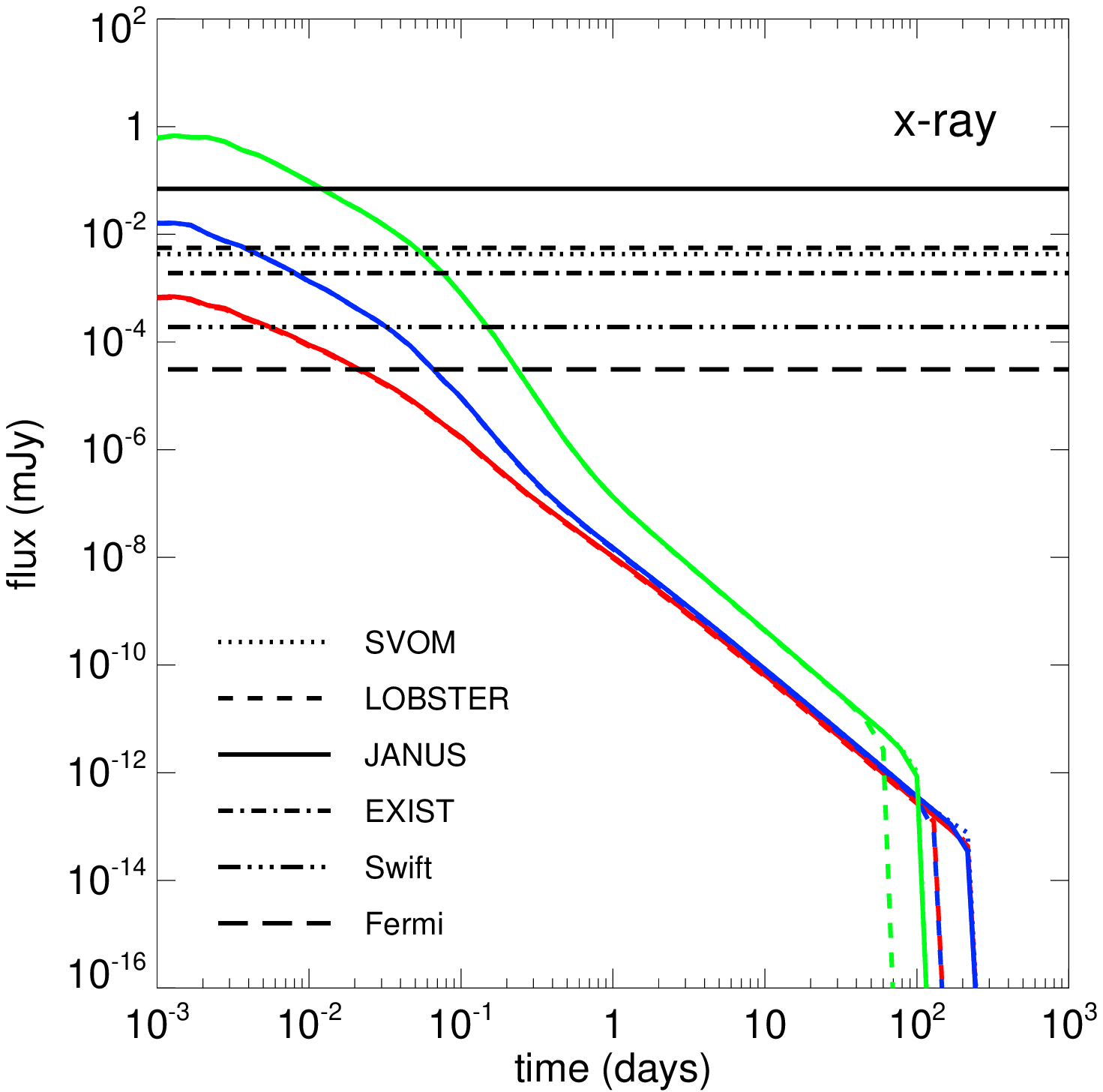,width=0.85\linewidth,clip=}  \\
\epsfig{file=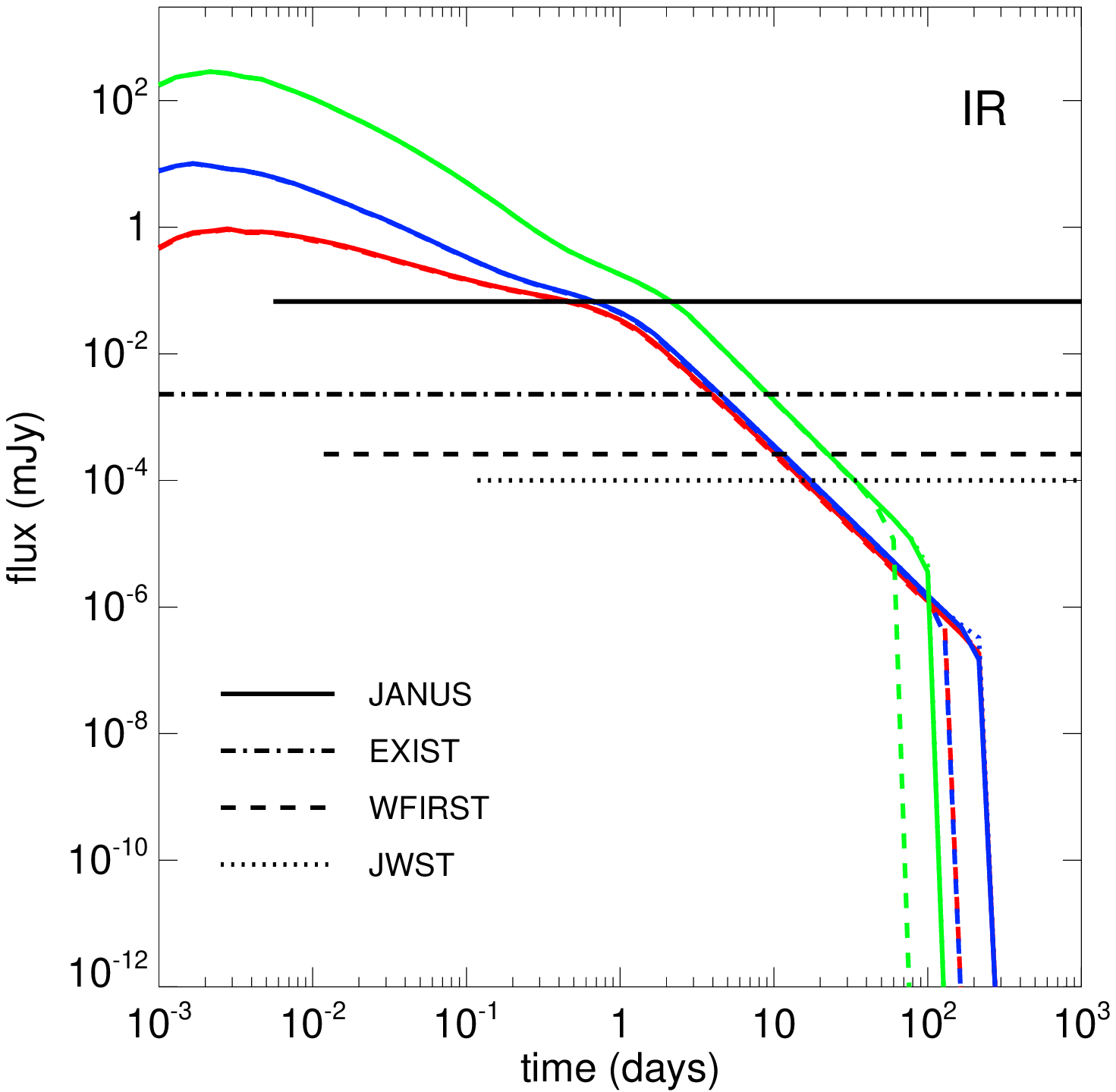,width=0.85\linewidth,clip=}  \\ 
\epsfig{file=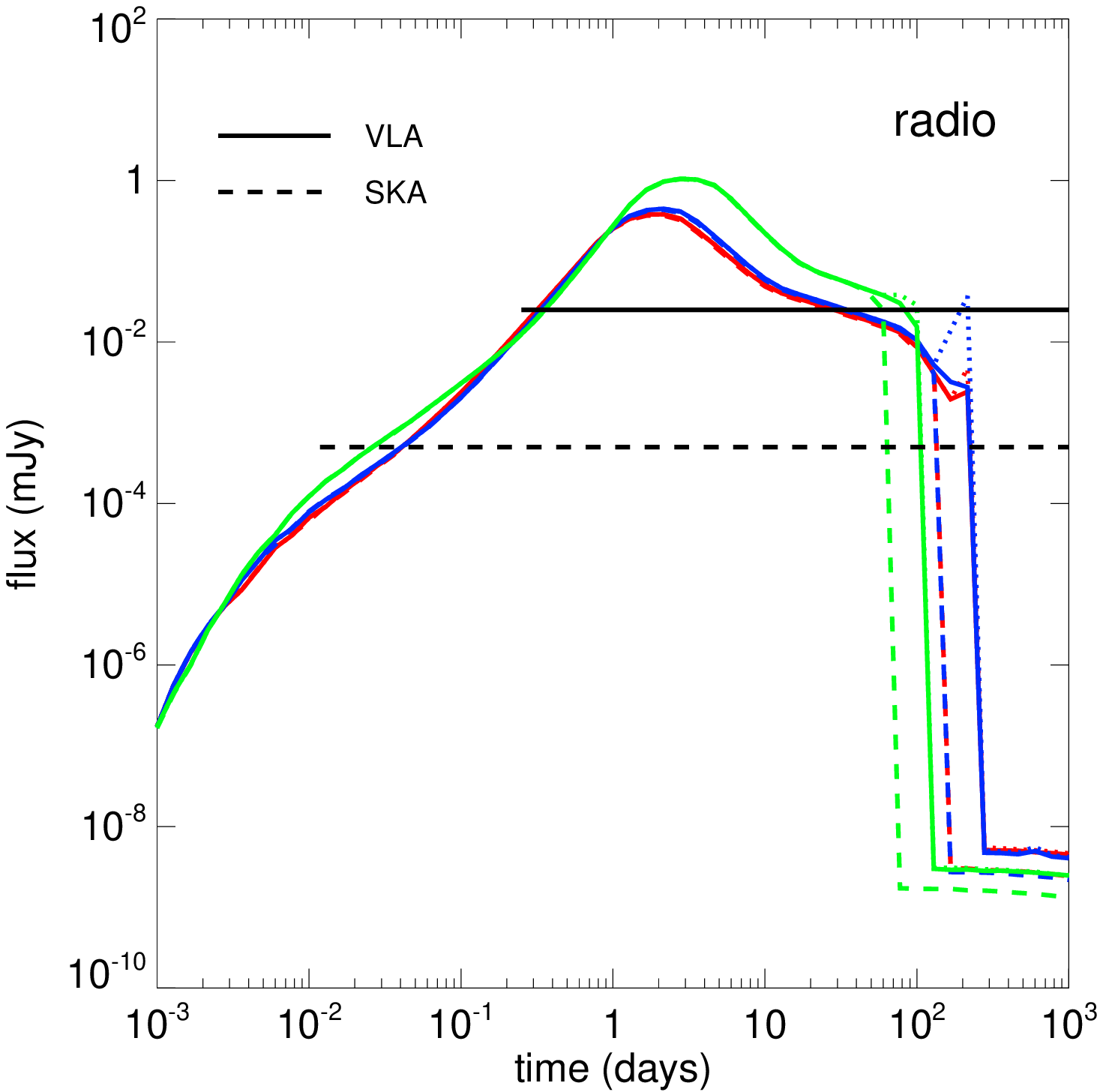,width=0.85\linewidth,clip=}  
\end{tabular}
\end{center}
\caption{Afterglow light curves for Pop III GRBs in 5 \Ms\ shells at 0.2 pc
ejected in binary mergers.  $E_{\mathrm{iso},\gamma} =$ 10$^{51}$ erg 
(red), 10$^{52}$ erg (blue), and 10$^{53}$ erg (green). Top: X-ray (82.7 
keV); center: NIR (3.0 $\mu$m); bottom: radio (5 GHz). Solid, dotted and 
dashed lines are for shells expanding in ambient densities $n =$ 0.1, 1.0, 
and 10 cm$^{-1}$, respectively.  All times are in the earth frame.}
\label{fig:BM}
\end{figure}

We plot afterglow light curves for Pop III GRBs in dense shells ejected in 
binary mergers in Figure~\ref{fig:BM}.  The hydrogen shell has a mass of 5 
\Ms\ and is driven to a radius of $\sim$ 0.2 pc by a 10$^{-5}$ \Ms\ yr$^{-1}$ 
wind in \HII\ region densities of 0.1, 1, and 10 cm$^{-3}$.  The jet initially 
breaks out into the wind and then reaches the termination shock at about 
10 days, where the wind has piled up at the inner surface of the shell.  By 
this time the jet has decelerated to $\Gamma \sim 10$.  A mildly relativistic 
reverse shock forms at the interface of the free-streaming and shocked 
winds and then propagates back into the jet.  At the same time, a mildly 
relativistic forward shock advances into the shocked wind.  As with the He 
merger shell, a series of forward and reverse shock pairs form at the 
interface between the free-streaming and shocked winds until the leading 
forward shock has fully advanced into the piled-up wind at the inner surface 
of the shell.  The jet, now only mildly relativistic, reaches the shell itself at 
several hundred to 1000 days, at which time it becomes non-relativistic.  A 
series of shock pairs are again created as the jet enters the shell as in the 
He merger case.  The jet eventually breaks out of the hydrogen shell and 
into the \HII\ region, but only after several thousand days.  

As shown in Figure~\ref{fig:BM}, the afterglow light curves are similar to those
in stellar winds out to 20 - 50 days.  Although the jet collides with the shocked 
wind after $\sim5$ days, the effect on the light curve is minimal. Only when the 
jet reaches the hydrogen shell at 20-50 days does the light curve deviate from 
that of a simple wind.  In the X-ray band, the afterglow falls below the detection 
limits of current and proposed instruments before the jet collides with the shell.  
In the IR, the flux is only $10^{-5} - 10^{-4}$ mJy when the shell is reached, 
making the transition into the shell barely detectable with {\it JWST} or perhaps 
WFIRST, and then only for the most energetic bursts.  Likewise, at 5 GHz the 
afterglow flux is only $\sim10^{-2}$ mJy when the jet crashes into the shell. The 
VLA may barely be able to detect the sudden drop in flux when the jet collides 
with the shell, but the SKA will detect it. Note that in some cases the drop in flux 
will be preceded by a flare in the radio.  Pop III GRBs due to binary mergers 
would look like events in simple winds, even when the shell is relatively close to
the progenitor because the X-ray and NIR fluxes would fall below detectability
before plummeting when the jet reaches the shell.  However, they could still be
distinguished from GRBs in winds by followup observations in the radio at later
times, which could capture the abrupt drop in flux.  Even the least energetic Pop
III GRB from a binary merger will be visible in the NIR to all the instruments in
Table \ref{tab:instr} for at least a day.

Our models show that what primarily governs the light curves of Pop III GRBs 
due to He and binary mergers for a given energy are the mass loss rates after 
the ejection (and hence the density of the wind envelope) and the time between 
the ejection and the burst.  The mass of the shell does not matter, as long as it 
can fully quench the jet.  

\subsection{Flares}

\begin{figure}
\plotone{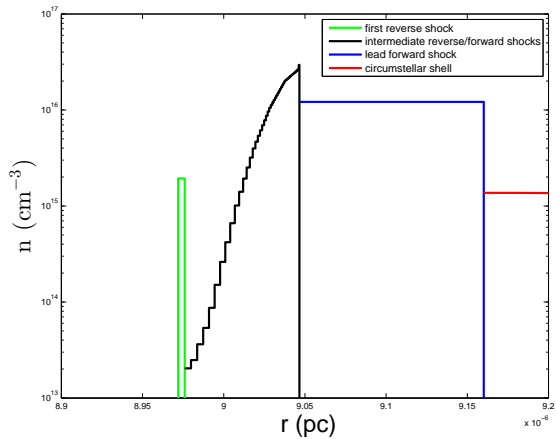} 
\caption{Density structure of the jet after it has collided with the dense shell.  
The red region is the local density of the shell at the leading edge of the jet.  
The blue region marks the leading forward shock created by the interaction 
of the jet with the most recent density jump.  This shock advances into the 
shell ahead of the rest of the jet.  The black region shows the series of 
forward and reverse shock pairs that were formed as the jet collides with the 
shell.  The spike (green) at the rear of the jet is the ambient material swept 
up before the jet reaches the shell. The reverse shock that formed when the 
jet with the shell has backstepped past the initial forward shock. \label{fig:jet}}
\vspace{0.1in}
\label{fig:jet}
\end{figure}

At least half of GRBs with observed afterglows exhibit some sort of flaring in 
their light curves \citep[see, e.g., ][]{rom12}.  The leading contender for the 
origin of these flares is late-time central engine activity.  The collision of the 
relativistic jet with strong density features has also been suggested as the 
cause of some flares, like those in our He and binary merger models (M12).  
But some have questioned if flares really occur when a GRB jet collides with 
a dense shell \citep{gatEA13}.  When the jet encounters the shell, a pair of 
forward and reverse shocks is formed, as noted earlier.  The new forward 
shock continues into the shell while the reverse shock steps back through 
the old forward shock in the frame of the shock.  Material swept up by the 
new forward shock does not mix with material that was swept up prior to the 
collision because of the contact discontinuity between the reverse shock and 
the new forward shock.  The M12 afterglow model, which employs the swept
mass approximation to model jet evolution, may therefore overestimate the 
mass that radiates at the post-collision Lorentz factor. \citet{gatEA13} argue 
that this could lead to overestimates of the flux at the interface between the 
shocked wind and the hydrogen shell, and hence a flare when none should 
exist.

The \citet{gatEA13} model accounts for the radial structure of the jet when 
modeling the passage of the reverse shock through the original forward 
shock during a collision with a density jump. But their model does not resolve 
the density structure at the inner surface of a shell.  It instead treats the inner 
surface as a single density jump of factor $a = 10^5$.  The radial structure of 
a jet that has collided with a single jump $a$ is not the same as for a jet that 
has traversed a series of smaller jumps that sum up to $a$.  Indeed, when we 
apply our model to the jets and density profiles in \citet{gatEA13}, it does not 
produce any flares either.

Figure~\ref{fig:jet} shows the radial structure of a jet after it has crashed into 
the 5 \Ms\ He merger shell in our study.  The jet does not instantaneously 
transition into the shell, but instead encounters a series of density jumps over 
$\sim 2$ hours.  Consequently, the structure of the jet becomes highly complex, 
with the vast majority of the swept-up mass being confined to a narrow region 
at the leading edge of the jet.  The potential smearing out of flares due to the 
radial extent of the jet is therefore far less than that predicted by the \citet{
gatEA13} model.  A relativistic blast wave colliding with this massive shell will 
thus produce a flare, even though the reverse shock and the curvature of the 
blast wave may reduce its amplitude.   

\section{Conclusion}

Pop III GRBs, together with other primordial SNe \citep{wet08a,fwf10,kasen11,
tomin11,hum12,tet12,mw12,pan12a,fet12,moriya12,wet12b,wet12e,wet12d,
wet12a,wet12c,jet13a,wet13a,jlj12a,wet13b,ds13,ds14,wet13c,wet13e,
chen14a,chen14b,chen14c}, will soon directly probe the properties of the first 
stars in the universe in X-rays, the NIR, and radio. GRBs will be easily 
distinguished from other SNe at this epoch by their prompt gamma emission, 
afterglow X-rays, relatively short-lived NIR profiles, and their appearance at 
later times in the radio.  Their environments, and hence their progenitors, can 
be deduced to some degree from the structures of their light curves.  The 
detection of Pop III GRB afterglows is crucial to pinpointing their redshifts, 
since at $z \sim$ 20 there is no host galaxy from which to infer a redshift, and 
even at $z \sim$ 10 - 15 the protogalaxy would probably be too dim to be 
observed.  Our algorithm produces afterglow light curves that are in good 
agreement with past work but goes well beyond them, giving realistic afterglow 
profiles for GRBs due to He mergers and binary mergers, which have not been 
modeled until now.

Bursts in winds are easily separated from GRBs in \HII\ regions by their much 
brighter NIR and X-ray fluxes at early times.  They also peak in the radio 
at much later times, 1- 30 days versus at most $\sim$ 1 day.  GRBs from He 
mergers have unique light curves that are characterized by prominent flares 
and abrupt quenching at less than a day across all bands. GRBs from binary 
mergers have light curves that are similar to those in winds except at later 
times, when their flux is abruptly cut off by the collision of the jet with the shell.  
This feature will generally only be visible in the radio because the X-ray and 
NIR fluxes will have fallen below the detection threshold of all current and 
proposed missions before they are cut off.  The sudden drop in radio flux in 
this type of event is sometimes preceded by a flare.  There is some 
degeneracy in energy and local environment with light curves across 
progenitor type that can complicate their classification, but in many cases it 
will still be possible to identify their progenitors.  

Although Pop III GRBs at $z \sim$ 20 can be detected by {\it Swift} and VLA
today, identifying them as being primordial can be problematic because:

\begin{enumerate}
\item{a Pop III GRB cannot necessarily be distinguished from an event at low 
redshift from the duration of its prompt gamma emission, which would be 
cosmologically dilated in time by factors of up to 20.  These times can still fall 
well within the wide range of prompt emission times found for GRBs in the local 
universe today.}

\item{even if prompt emission is suspected to be from a primordial event, its 
NIR afterglow, which could confirm its redshift, would not be visible to most 
facilities today, and may be too transient to be captured by those that can
detect them.}

\item{if prompt gamma emission is suspected to be from a Pop III GRB, it may 
not be accompanied by a radio afterglow that is strong enough to determine 
its redshift (only $\sim$ 10\% of GRBs have measurable radio signatures).}

\end{enumerate}

\noindent Some of these problems would persist even with the next generation 
of NIR and radio observatories.  For example, if {\it Swift} detected a burst that 
was thought to be Pop III, its NIR signature would disappear before {\it JWST} 
or a ground-based extremely large telescope could be tasked to observe it if it 
is a He merger event. Indeed, in spite of its extreme sensitivity in the NIR, {\it 
JWST} will probably not be used to follow up on future GRBs because doing 
so would rapidly deplete its limited fuel supply. GRB rates at $z \sim$ 20 would 
also be so low that it is highly unlikely that the narrow {\it JWST} fields would 
encounter one in routine surveys over its mission lifetime.

Future successors to {\it Swift} such as EXIST and JANUS will be the best 
equipped to hunt for Pop III GRBs for two reasons:

\begin{enumerate}

\item{they will be all-sky X-ray and NIR missions whose wide fields will partly 
compensate for the low GRB rates at $z \gtrsim$ 10 predicted by some studies.}

\item{their onboard telescopes will be able to measure the NIR flux of a Pop III 
GRB from the moment of the burst, thereby determining its redshift and, in 
some cases, progenitor type.}

\end{enumerate}

\noindent For example, GRBs due to He mergers would, in all likelihood, only 
be detected by missions with onboard NIR telescopes.  Our models show that
NIR instruments with the sensitivity of the EXIST IRT would detect even the 
least energetic events at $z \sim$ 20 but that the JANUS NIRT in some cases
would not.  

Our simulations also show that Pop III GRBs would appear in future NIR and 
radio surveys, regardless of prompt gamma emission.  Radio afterglows at $z 
\sim$ 20 would appear in ground-based campaigns by VLA and SKA with an 
appropriate cadence.  These events could be easily be separated from the 
radio signatures of Pop III SNe, which evolve on much longer timescales \citep{
mw12}.  Further studies are needed to determine if Pop III GRBs and SNe can
be distinguished from the many foreground sources of synchrotron emission 
that could contaminate their signal.  Equally important, most Pop III GRBs are
bright enough to appear in all-sky surveys at $z \sim$ 20 by WFIRST and the 
Wide-Field Imaging Surveyor for High Redshift (WISH).  The large search
areas of these missions could compensate for the low event numbers at such 
redshifts.  It should be noted that although Pop III GRBs may be more visible in 
the NIR than in the radio, radio facilities exist that are capable of seeing them 
now.

Hypernovae (HNe), highly energetic Type Ib/c SNe, may be associated with 
some Pop III GRBs.  But new studies show that if the GRB is visible, it will 
completely outshine the HN \citep{smidt13a}.  If not, the HN itself will only be 
visible to {\it JWST} out to $z \sim$ 10 - 15, the era of first galaxy formation, 
and to WFIRST at $z \sim$ 4 - 6, the end of reionization.  In the upper limit 
that there is a HN with every GRB, the HN rate would be $\sim$ 100 times 
the detected GRB rate at a given redshift for typical opening angles for the jet.  
Even at these rates, it is unlikely that the HN would be found by {\it JWST} or 
the next generation of ELTs without its GRB first being detected, and they lie 
beyond the reach of all-sky NIR missions at $z \gtrsim$ 6, which might 
otherwise have detected them because of their wide fields.  But if the GRB is
detected at $z \lesssim$ 15, followup observations might reveal a HN.

Our suite of Pop III GRB light curves is not comprehensive.  First, they do not 
account for multidimensional structures in the vicinity of the burst.  Massive
shells ejected in mergers are prone to dynamical instabilities that could fracture
them into clumps, and emission from a relativistic jet crashing into a clump 
would be quite different than if it were piercing a crack in the shell. Second, as
noted earlier, the ambient medium of the burst could be intermediate to the 
four canonical cases considered here.  Clumps might exist in \HII\ regions in
the absence of winds, and GRB jets could break out along an axis perpendicular 
to the plane of a toroidal ejection.  Finally, new models such as high-resolution
special-relativistic hydrodynamics and particle-in-cell (PIC) simulations are 
needed to better probe the microphysical processes that contribute to the 
afterglow flux.  Future multidimensional simulations in this vein can also 
address the imprint of realistic circumburst structures on the afterglows of Pop 
III GRBs.

\acknowledgments

RM was supported by LANL IGPP grant 10-150.  D.J.W. was supported by 
the European Research Council under the European Community's Seventh 
Framework Programme (FP7/2007 - 2013) via the ERC Advanced Grant 
"STARLIGHT:  Formation of the First Stars" (project number 339177).  Work 
at LANL was done under the auspices of the National Nuclear Security 
Administration of the U.S. Department of Energy at Los Alamos National 
Laboratory under Contract No. DE-AC52-06NA25396.  All ZEUS-MP 
simulations were performed with allocations from Institutional Computing (IC) 
on the Pinto cluster at LANL.  

\bibliographystyle{apj}
\bibliography{refs}

\begin{thebibliography}{131}
\expandafter\ifx\csname natexlab\endcsname\relax\def\natexlab#1{#1}\fi

\bibitem[{{Abel} {et~al.}(2007){Abel}, {Wise}, \& {Bryan}}]{awb07}
{Abel}, T., {Wise}, J.~H., \& {Bryan}, G.~L. 2007, \apjl, 659, L87

\bibitem[{{Baraffe} {et~al.}(2001){Baraffe}, {Heger}, \& {Woosley}}]{baraffe01}
{Baraffe}, I., {Heger}, A., \& {Woosley}, S.~E. 2001, \apj, 550, 890

\bibitem[{{Barkana} \& {Loeb}(2004)}]{brk04}
{Barkana}, R. \& {Loeb}, A. 2004, \apj, 601, 64

\bibitem[{{Beers} \& {Christlieb}(2005)}]{bc05}
{Beers}, T.~C. \& {Christlieb}, N. 2005, \araa, 43, 531

\bibitem[{{Blandford} \& {McKee}(1976)}]{blandfordMcKee76}
{Blandford}, R.~D. \& {McKee}, C.~F. 1976, Physics of Fluids, 19, 1130

\bibitem[{{Bromm} \& {Loeb}(2002)}]{bl02}
{Bromm}, V. \& {Loeb}, A. 2002, \apj, 575, 111

\bibitem[{{Bromm} \& {Loeb}(2006{\natexlab{a}})}]{bl06b}
{Bromm}, V. \& {Loeb}, A. 2006{\natexlab{a}}, in American Institute of Physics
  Conference Series, Vol. 836, Gamma-Ray Bursts in the Swift Era, ed. S.~S.
  {Holt}, N.~{Gehrels}, \& J.~A. {Nousek}, 503--512

\bibitem[{{Bromm} \& {Loeb}(2006{\natexlab{b}})}]{bl06a}
---. 2006{\natexlab{b}}, \apj, 642, 382

\bibitem[{{Bromm} {et~al.}(2009){Bromm}, {Yoshida}, {Hernquist}, \&
  {McKee}}]{fsg09}
{Bromm}, V., {Yoshida}, N., {Hernquist}, L., \& {McKee}, C.~F. 2009, \nat, 459,
  49

\bibitem[{{Burrows} {et~al.}(2010){Burrows}, {Roming}, {Fox}, {Herter},
  {Falcone}, {Bil{\'e}n}, {Nousek}, \& {Kennea}}]{Burrows10}
{Burrows}, D.~N., {Roming}, P.~W.~A., {Fox}, D.~B., {Herter}, T.~L., {Falcone},
  A., {Bil{\'e}n}, S., {Nousek}, J.~A., \& {Kennea}, J.~A. 2010, in Presented
  at the Society of Photo-Optical Instrumentation Engineers (SPIE) Conference,
  Vol. 7732, Society of Photo-Optical Instrumentation Engineers (SPIE)
  Conference Series

\bibitem[{{Caffau} {et~al.}(2012){Caffau}, {Bonifacio}, {Fran{\c c}ois},
  {Spite}, {Spite}, {Zaggia}, {Ludwig}, {Steffen}, {Mashonkina}, {Monaco},
  {Sbordone}, {Molaro}, {Cayrel}, {Plez}, {Hill}, {Hammer}, \&
  {Randich}}]{caffau12}
{Caffau}, E., {Bonifacio}, P., {Fran{\c c}ois}, P., {Spite}, M., {Spite}, F.,
  {Zaggia}, S., {Ludwig}, H.-G., {Steffen}, M., {Mashonkina}, L., {Monaco}, L.,
  {Sbordone}, L., {Molaro}, P., {Cayrel}, R., {Plez}, B., {Hill}, V., {Hammer},
  F., \& {Randich}, S. 2012, \aap, 542, A51

\bibitem[{{Chen} {et~al.}(2014{\natexlab{a}}){Chen}, {Heger}, {Woosley},
  {Almgren}, \& {Whalen}}]{chen14c}
{Chen}, K.-J., {Heger}, A., {Woosley}, S., {Almgren}, A., \& {Whalen}, D.
  2014{\natexlab{a}}, arXiv:1402.5960

\bibitem[{{Chen} {et~al.}(2014{\natexlab{b}}){Chen}, {Heger}, {Woosley},
  {Almgren}, {Whalen}, \& {Johnson}}]{chen14b}
{Chen}, K.-J., {Heger}, A., {Woosley}, S., {Almgren}, A., {Whalen}, D., \&
  {Johnson}, J. 2014{\natexlab{b}}, arXiv:1402.4777

\bibitem[{{Chen} {et~al.}(2014{\natexlab{c}}){Chen}, {Woosley}, {Heger},
  {Almgren}, \& {Whalen}}]{chen14a}
{Chen}, K.-J., {Woosley}, S., {Heger}, A., {Almgren}, A., \& {Whalen}, D.
  2014{\natexlab{c}}, arXiv:1402.4134

\bibitem[{{Ciardi} \& {Loeb}(2000)}]{cl00}
{Ciardi}, B. \& {Loeb}, A. 2000, \apj, 540, 687

\bibitem[{{Clark} {et~al.}(2011){Clark}, {Glover}, {Smith}, {Greif}, {Klessen},
  \& {Bromm}}]{clark11}
{Clark}, P.~C., {Glover}, S.~C.~O., {Smith}, R.~J., {Greif}, T.~H., {Klessen},
  R.~S., \& {Bromm}, V. 2011, Science, 331, 1040

\bibitem[{{Cucchiara} {et~al.}(2011){Cucchiara}, {Levan}, {Fox}, {Tanvir},
  {Ukwatta}, {Berger}, {Kr{\"u}hler}, {K{\"u}pc{\"u} Yolda{\c s}}, {Wu},
  {Toma}, {Greiner}, {Olivares}, {Rowlinson}, {Amati}, {Sakamoto}, {Roth},
  {Stephens}, {Fritz}, {Fynbo}, {Hjorth}, {Malesani}, {Jakobsson}, {Wiersema},
  {O'Brien}, {Soderberg}, {Foley}, {Fruchter}, {Rhoads}, {Rutledge}, {Schmidt},
  {Dopita}, {Podsiadlowski}, {Willingale}, {Wolf}, {Kulkarni}, \&
  {D'Avanzo}}]{cucc11}
{Cucchiara}, A., {Levan}, A.~J., {Fox}, D.~B., {Tanvir}, N.~R., {Ukwatta},
  T.~N., {Berger}, E., {Kr{\"u}hler}, T., {K{\"u}pc{\"u} Yolda{\c s}}, A.,
  {Wu}, X.~F., {Toma}, K., {Greiner}, J., {Olivares}, F.~E., {Rowlinson}, A.,
  {Amati}, L., {Sakamoto}, T., {Roth}, K., {Stephens}, A., {Fritz}, A.,
  {Fynbo}, J.~P.~U., {Hjorth}, J., {Malesani}, D., {Jakobsson}, P., {Wiersema},
  K., {O'Brien}, P.~T., {Soderberg}, A.~M., {Foley}, R.~J., {Fruchter}, A.~S.,
  {Rhoads}, J., {Rutledge}, R.~E., {Schmidt}, B.~P., {Dopita}, M.~A.,
  {Podsiadlowski}, P., {Willingale}, R., {Wolf}, C., {Kulkarni}, S.~R., \&
  {D'Avanzo}, P. 2011, \apj, 736, 7

\bibitem[{{Dai} {et~al.}(1999){Dai}, {Huang}, \& {Lu}}]{daiEA99}
{Dai}, Z.~G., {Huang}, Y.~F., \& {Lu}, T. 1999, \apj, 520, 634

\bibitem[{Dai \& Lu(2002)}]{daiLu02}
Dai, Z.~G. \& Lu, T. 2002, \apj, 565, L87

\bibitem[{{de Souza} {et~al.}(2013){de Souza}, {Ishida}, {Johnson}, {Whalen},
  \& {Mesinger}}]{ds13}
{de Souza}, R.~S., {Ishida}, E.~E.~O., {Johnson}, J.~L., {Whalen}, D.~J., \&
  {Mesinger}, A. 2013, \mnras, 436, 1555

\bibitem[{{de Souza} {et~al.}(2014){de Souza}, {Ishida}, {Whalen}, {Johnson},
  \& {Ferrara}}]{ds14}
{de Souza}, R.~S., {Ishida}, E.~E.~O., {Whalen}, D.~J., {Johnson}, J., \&
  {Ferrara}, A. 2014, arXiv:1401.2995

\bibitem[{{de Souza} {et~al.}(2011){de Souza}, {Yoshida}, \& {Ioka}}]{ds11}
{de Souza}, R.~S., {Yoshida}, N., \& {Ioka}, K. 2011, \aap, 533, A32

\bibitem[{{Falcone} {et~al.}(2009){Falcone}, {Burrows}, {Barthelmy}, {Chang},
  {Fredley}, {Kelly}, {Klar}, {Palmer}, {Persyn}, {Reichard}, {Roming},
  {Seifert}, {Smith}, {Wood}, \& {Zugger}}]{falconeEA09}
{Falcone}, A.~D., {Burrows}, D.~N., {Barthelmy}, S., {Chang}, W., {Fredley},
  J., {Kelly}, M., {Klar}, R., {Palmer}, D., {Persyn}, S., {Reichard}, K.,
  {Roming}, P., {Seifert}, E., {Smith}, R.~W.~M., {Wood}, P., \& {Zugger}, M.
  2009, in Society of Photo-Optical Instrumentation Engineers (SPIE) Conference
  Series, Vol. 7435, Society of Photo-Optical Instrumentation Engineers (SPIE)
  Conference Series

\bibitem[{{Fenimore} {et~al.}(1996){Fenimore}, {Madras}, \&
  {Nayakshin}}]{fenimoreEA96}
{Fenimore}, E.~E., {Madras}, C.~D., \& {Nayakshin}, S. 1996, \apj, 473, 998

\bibitem[{{Frebel} {et~al.}(2005){Frebel}, {Aoki}, {Christlieb}, {Ando},
  {Asplund}, {Barklem}, {Beers}, {Eriksson}, {Fechner}, {Fujimoto}, {Honda},
  {Kajino}, {Minezaki}, {Nomoto}, {Norris}, {Ryan}, {Takada-Hidai},
  {Tsangarides}, \& {Yoshii}}]{fet05}
{Frebel}, A., {Aoki}, W., {Christlieb}, N., {Ando}, H., {Asplund}, M.,
  {Barklem}, P.~S., {Beers}, T.~C., {Eriksson}, K., {Fechner}, C., {Fujimoto},
  M.~Y., {Honda}, S., {Kajino}, T., {Minezaki}, T., {Nomoto}, K., {Norris},
  J.~E., {Ryan}, S.~G., {Takada-Hidai}, M., {Tsangarides}, S., \& {Yoshii}, Y.
  2005, \nat, 434, 871

\bibitem[{{Frey} {et~al.}(2013){Frey}, {Even}, {Whalen}, {Fryer}, {Hungerford},
  {Fontes}, \& {Colgan}}]{fet12}
{Frey}, L.~H., {Even}, W., {Whalen}, D.~J., {Fryer}, C.~L., {Hungerford},
  A.~L., {Fontes}, C.~J., \& {Colgan}, J. 2013, \apjs, 204, 16

\bibitem[{{Fryer} {et~al.}(2013){Fryer}, {Belczynski}, {Berger}, {Th{\"o}ne},
  {Ellinger}, \& {Bulik}}]{fb13}
{Fryer}, C.~L., {Belczynski}, K., {Berger}, E., {Th{\"o}ne}, C., {Ellinger},
  C., \& {Bulik}, T. 2013, \apj, 764, 181

\bibitem[{{Fryer} \& {Heger}(2005)}]{fh05}
{Fryer}, C.~L. \& {Heger}, A. 2005, \apj, 623, 302

\bibitem[{{Fryer} \& {Heger}(2011)}]{fh11}
---. 2011, Astronomische Nachrichten, 332, 408

\bibitem[{{Fryer} {et~al.}(2007){Fryer}, {Mazzali}, {Prochaska}, {Cappellaro},
  {Panaitescu}, {Berger}, {van Putten}, {van den Heuvel}, {Young},
  {Hungerford}, {Rockefeller}, {Yoon}, {Podsiadlowski}, {Nomoto}, {Chevalier},
  {Schmidt}, \& {Kulkarni}}]{pasp07}
{Fryer}, C.~L., {Mazzali}, P.~A., {Prochaska}, J., {Cappellaro}, E.,
  {Panaitescu}, A., {Berger}, E., {van Putten}, M., {van den Heuvel}, E.~P.~J.,
  {Young}, P., {Hungerford}, A., {Rockefeller}, G., {Yoon}, S.-C.,
  {Podsiadlowski}, P., {Nomoto}, K., {Chevalier}, R., {Schmidt}, B., \&
  {Kulkarni}, S. 2007, \pasp, 119, 1211

\bibitem[{{Fryer} {et~al.}(2010){Fryer}, {Whalen}, \& {Frey}}]{fwf10}
{Fryer}, C.~L., {Whalen}, D.~J., \& {Frey}, L. 2010, in American Institute of
  Physics Conference Series, Vol. 1294, American Institute of Physics
  Conference Series, ed. D.~J. {Whalen}, V.~{Bromm}, \& N.~{Yoshida}, 70--75

\bibitem[{{Fryer} \& {Woosley}(1998)}]{fw98}
{Fryer}, C.~L. \& {Woosley}, S.~E. 1998, \apjl, 502, L9

\bibitem[{{Fryer} {et~al.}(1999){Fryer}, {Woosley}, \& {Hartmann}}]{fwh99}
{Fryer}, C.~L., {Woosley}, S.~E., \& {Hartmann}, D.~H. 1999, \apj, 526, 152

\bibitem[{{Fryer} {et~al.}(2001){Fryer}, {Woosley}, \& {Heger}}]{fwh01}
{Fryer}, C.~L., {Woosley}, S.~E., \& {Heger}, A. 2001, \apj, 550, 372

\bibitem[{{Gat} {et~al.}(2013){Gat}, {van Eerten}, \& {MacFadyen}}]{gatEA13}
{Gat}, I., {van Eerten}, H., \& {MacFadyen}, A. 2013, \apj, 773, 2

\bibitem[{{Glover}(2013)}]{glov12}
{Glover}, S. 2013, in Astrophysics and Space Science Library, Vol. 396,
  Astrophysics and Space Science Library, ed. T.~{Wiklind}, B.~{Mobasher}, \&
  V.~{Bromm}, 103

\bibitem[{{Gorenstein}(2011)}]{goren11}
{Gorenstein}, P. 2011, in Society of Photo-Optical Instrumentation Engineers
  (SPIE) Conference Series, Vol. 8147, Society of Photo-Optical Instrumentation
  Engineers (SPIE) Conference Series

\bibitem[{{Gou} {et~al.}(2004){Gou}, {M{\'e}sz{\'a}ros}, {Abel}, \&
  {Zhang}}]{gou04}
{Gou}, L.~J., {M{\'e}sz{\'a}ros}, P., {Abel}, T., \& {Zhang}, B. 2004, \apj,
  604, 508

\bibitem[{{Greif} {et~al.}(2012){Greif}, {Bromm}, {Clark}, {Glover}, {Smith},
  {Klessen}, {Yoshida}, \& {Springel}}]{get12}
{Greif}, T.~H., {Bromm}, V., {Clark}, P.~C., {Glover}, S.~C.~O., {Smith},
  R.~J., {Klessen}, R.~S., {Yoshida}, N., \& {Springel}, V. 2012, \mnras, 424,
  399

\bibitem[{{Greiner} {et~al.}(2009){Greiner}, {Kr{\"u}hler}, {Fynbo}, {Rossi},
  {Schwarz}, {Klose}, {Savaglio}, {Tanvir}, {McBreen}, {Totani}, {Zhang}, {Wu},
  {Watson}, {Barthelmy}, {Beardmore}, {Ferrero}, {Gehrels}, {Kann}, {Kawai},
  {Yolda{\c s}}, {M{\'e}sz{\'a}ros}, {Milvang-Jensen}, {Oates}, {Pierini},
  {Schady}, {Toma}, {Vreeswijk}, {Yolda{\c s}}, {Zhang}, {Afonso}, {Aoki},
  {Burrows}, {Clemens}, {Filgas}, {Haiman}, {Hartmann}, {Hasinger}, {Hjorth},
  {Jehin}, {Levan}, {Liang}, {Malesani}, {Pyo}, {Schulze}, {Szokoly}, {Terada},
  \& {Wiersema}}]{grein09}
{Greiner}, J., {Kr{\"u}hler}, T., {Fynbo}, J.~P.~U., {Rossi}, A., {Schwarz},
  R., {Klose}, S., {Savaglio}, S., {Tanvir}, N.~R., {McBreen}, S., {Totani},
  T., {Zhang}, B.~B., {Wu}, X.~F., {Watson}, D., {Barthelmy}, S.~D.,
  {Beardmore}, A.~P., {Ferrero}, P., {Gehrels}, N., {Kann}, D.~A., {Kawai}, N.,
  {Yolda{\c s}}, A.~K., {M{\'e}sz{\'a}ros}, P., {Milvang-Jensen}, B., {Oates},
  S.~R., {Pierini}, D., {Schady}, P., {Toma}, K., {Vreeswijk}, P.~M., {Yolda{\c
  s}}, A., {Zhang}, B., {Afonso}, P., {Aoki}, K., {Burrows}, D.~N., {Clemens},
  C., {Filgas}, R., {Haiman}, Z., {Hartmann}, D.~H., {Hasinger}, G., {Hjorth},
  J., {Jehin}, E., {Levan}, A.~J., {Liang}, E.~W., {Malesani}, D., {Pyo},
  T.-S., {Schulze}, S., {Szokoly}, G., {Terada}, K., \& {Wiersema}, K. 2009,
  \apj, 693, 1610

\bibitem[{{Greiner} {et~al.}(2012){Greiner}, {Mannheim}, {Aharonian}, {Ajello},
  {Balasz}, {Barbiellini}, {Bellazzini}, {Bishop}, {Bisnovatij-Kogan}, {Boggs},
  {Bykov}, {DiCocco}, {Diehl}, {Els{\"a}sser}, {Foley}, {Fransson}, {Gehrels},
  {Hanlon}, {Hartmann}, {Hermsen}, {Hillebrandt}, {Hudec}, {Iyudin}, {Jose},
  {Kadler}, {Kanbach}, {Klamra}, {Kiener}, {Klose}, {Kreykenbohm}, {Kuiper},
  {Kylafis}, {Labanti}, {Langanke}, {Langer}, {Larsson}, {Leibundgut}, {Laux},
  {Longo}, {Maeda}, {Marcinkowski}, {Marisaldi}, {McBreen}, {McBreen},
  {Meszaros}, {Nomoto}, {Pearce}, {Peer}, {Pian}, {Prantzos}, {Raffelt},
  {Reimer}, {Rhode}, {Ryde}, {Schmidt}, {Silk}, {Shustov}, {Strong}, {Tanvir},
  {Thielemann}, {Tibolla}, {Tierney}, {Tr{\"u}mper}, {Varshalovich}, {Wilms},
  {Wrochna}, {Zdziarski}, \& {Zoglauer}}]{greinerEA12}
{Greiner}, J., {Mannheim}, K., {Aharonian}, F., {Ajello}, M., {Balasz}, L.~G.,
  {Barbiellini}, G., {Bellazzini}, R., {Bishop}, S., {Bisnovatij-Kogan}, G.~S.,
  {Boggs}, S., {Bykov}, A., {DiCocco}, G., {Diehl}, R., {Els{\"a}sser}, D.,
  {Foley}, S., {Fransson}, C., {Gehrels}, N., {Hanlon}, L., {Hartmann}, D.,
  {Hermsen}, W., {Hillebrandt}, W., {Hudec}, R., {Iyudin}, A., {Jose}, J.,
  {Kadler}, M., {Kanbach}, G., {Klamra}, W., {Kiener}, J., {Klose}, S.,
  {Kreykenbohm}, I., {Kuiper}, L.~M., {Kylafis}, N., {Labanti}, C., {Langanke},
  K., {Langer}, N., {Larsson}, S., {Leibundgut}, B., {Laux}, U., {Longo}, F.,
  {Maeda}, K., {Marcinkowski}, R., {Marisaldi}, M., {McBreen}, B., {McBreen},
  S., {Meszaros}, A., {Nomoto}, K., {Pearce}, M., {Peer}, A., {Pian}, E.,
  {Prantzos}, N., {Raffelt}, G., {Reimer}, O., {Rhode}, W., {Ryde}, F.,
  {Schmidt}, C., {Silk}, J., {Shustov}, B.~M., {Strong}, A., {Tanvir}, N.,
  {Thielemann}, F.-K., {Tibolla}, O., {Tierney}, D., {Tr{\"u}mper}, J.,
  {Varshalovich}, D.~A., {Wilms}, J., {Wrochna}, G., {Zdziarski}, A., \&
  {Zoglauer}, A. 2012, Experimental Astronomy, 34, 551

\bibitem[{{Grindlay}(2010)}]{grindlay10}
{Grindlay}, J.~E. 2010, in American Institute of Physics Conference Series,
  Vol. 1279, American Institute of Physics Conference Series, ed. N.~{Kawai} \&
  S.~{Nagataki}, 212--219

\bibitem[{{Heger} \& {Woosley}(2002)}]{hw02}
{Heger}, A. \& {Woosley}, S.~E. 2002, \apj, 567, 532

\bibitem[{{Hirano} {et~al.}(2013){Hirano}, {Hosokawa}, {Yoshida}, {Umeda},
  {Omukai}, {Chiaki}, \& {Yorke}}]{hir13}
{Hirano}, S., {Hosokawa}, T., {Yoshida}, N., {Umeda}, H., {Omukai}, K.,
  {Chiaki}, G., \& {Yorke}, H.~W. 2013, arXiv:1308.4456

\bibitem[{{Hong} {et~al.}(2009){Hong}, {Grindlay}, {Allen}, {Barthelmy},
  {Skinner}, \& {Gehrels}}]{hongEA09}
{Hong}, J., {Grindlay}, J.~E., {Allen}, B., {Barthelmy}, S.~D., {Skinner},
  G.~K., \& {Gehrels}, N. 2009, in Society of Photo-Optical Instrumentation
  Engineers (SPIE) Conference Series, Vol. 7435, Society of Photo-Optical
  Instrumentation Engineers (SPIE) Conference Series

\bibitem[{{Hosokawa} {et~al.}(2011){Hosokawa}, {Omukai}, {Yoshida}, \&
  {Yorke}}]{hos11}
{Hosokawa}, T., {Omukai}, K., {Yoshida}, N., \& {Yorke}, H.~W. 2011, Science,
  334, 1250

\bibitem[{Huang {et~al.}(1999)Huang, Dai, \& Lu}]{huangEA99}
Huang, Y., Dai, Z., \& Lu, T. 1999, \mnras, 309, 513

\bibitem[{{Huang} {et~al.}(2000){Huang}, {Gou}, {Dai}, \& {Lu}}]{huangEA00}
{Huang}, Y.~F., {Gou}, L.~J., {Dai}, Z.~G., \& {Lu}, T. 2000, \apj, 543, 90

\bibitem[{{Hummel} {et~al.}(2012){Hummel}, {Pawlik}, {Milosavljevi{\'c}}, \&
  {Bromm}}]{hum12}
{Hummel}, J.~A., {Pawlik}, A.~H., {Milosavljevi{\'c}}, M., \& {Bromm}, V. 2012,
  \apj, 755, 72

\bibitem[{{Inoue} {et~al.}(2007){Inoue}, {Omukai}, \& {Ciardi}}]{inoue07}
{Inoue}, S., {Omukai}, K., \& {Ciardi}, B. 2007, \mnras, 380, 1715

\bibitem[{{Ioka}(2003)}]{i03}
{Ioka}, K. 2003, \apjl, 598, L79

\bibitem[{{Ioka} \& {M{\'e}sz{\'a}ros}(2005)}]{im05}
{Ioka}, K. \& {M{\'e}sz{\'a}ros}, P. 2005, \apj, 619, 684

\bibitem[{{Jeon} {et~al.}(2012){Jeon}, {Pawlik}, {Greif}, {Glover}, {Bromm},
  {Milosavljevi{\'c}}, \& {Klessen}}]{jeon11}
{Jeon}, M., {Pawlik}, A.~H., {Greif}, T.~H., {Glover}, S.~C.~O., {Bromm}, V.,
  {Milosavljevi{\'c}}, M., \& {Klessen}, R.~S. 2012, \apj, 754, 34

\bibitem[{{Joggerst} {et~al.}(2010){Joggerst}, {Almgren}, {Bell}, {Heger},
  {Whalen}, \& {Woosley}}]{jet09b}
{Joggerst}, C.~C., {Almgren}, A., {Bell}, J., {Heger}, A., {Whalen}, D., \&
  {Woosley}, S.~E. 2010, \apj, 709, 11

\bibitem[{{Joggerst} \& {Whalen}(2011)}]{jw11}
{Joggerst}, C.~C. \& {Whalen}, D.~J. 2011, \apj, 728, 129

\bibitem[{{Johnson} {et~al.}(2013{\natexlab{a}}){Johnson}, {Whalen}, {Even},
  {Fryer}, {Heger}, {Smidt}, \& {Chen}}]{jet13a}
{Johnson}, J.~L., {Whalen}, D.~J., {Even}, W., {Fryer}, C.~L., {Heger}, A.,
  {Smidt}, J., \& {Chen}, K.-J. 2013{\natexlab{a}}, arXiv:1304.4601

\bibitem[{{Johnson} {et~al.}(2012){Johnson}, {Whalen}, {Fryer}, \&
  {Li}}]{jlj12a}
{Johnson}, J.~L., {Whalen}, D.~J., {Fryer}, C.~L., \& {Li}, H. 2012, \apj, 750,
  66

\bibitem[{{Johnson} {et~al.}(2013{\natexlab{b}}){Johnson}, {Whalen}, {Li}, \&
  {Holz}}]{jet13}
{Johnson}, J.~L., {Whalen}, D.~J., {Li}, H., \& {Holz}, D.~E.
  2013{\natexlab{b}}, \apj, 771, 116

\bibitem[{{Kasen} {et~al.}(2011){Kasen}, {Woosley}, \& {Heger}}]{kasen11}
{Kasen}, D., {Woosley}, S.~E., \& {Heger}, A. 2011, \apj, 734, 102

\bibitem[{{Kashiyama} {et~al.}(2013){Kashiyama}, {Nakauchi}, {Suwa}, {Yajima},
  \& {Nakamura}}]{kash13}
{Kashiyama}, K., {Nakauchi}, D., {Suwa}, Y., {Yajima}, H., \& {Nakamura}, T.
  2013, \apj, 770, 8

\bibitem[{{Keller} {et~al.}(2014){Keller}, {Bessell}, {Frebel}, {Casey},
  {Asplund}, {Jacobson}, {Lind}, {Norris}, {Yong}, {Heger}, {Magic}, {da
  Costa}, {Schmidt}, \& {Tisserand}}]{keller14}
{Keller}, S.~C., {Bessell}, M.~S., {Frebel}, A., {Casey}, A.~R., {Asplund}, M.,
  {Jacobson}, H.~R., {Lind}, K., {Norris}, J.~E., {Yong}, D., {Heger}, A.,
  {Magic}, Z., {da Costa}, G.~S., {Schmidt}, B.~P., \& {Tisserand}, P. 2014,
  \nat, 506, 463

\bibitem[{{Kitayama} {et~al.}(2004){Kitayama}, {Yoshida}, {Susa}, \&
  {Umemura}}]{ket04}
{Kitayama}, T., {Yoshida}, N., {Susa}, H., \& {Umemura}, M. 2004, \apj, 613,
  631

\bibitem[{{Lamb} \& {Reichart}(2000)}]{lr00}
{Lamb}, D.~Q. \& {Reichart}, D.~E. 2000, \apj, 536, 1

\bibitem[{{Laskar} {et~al.}(2014){Laskar}, {Berger}, {Tanvir}, {Zauderer},
  {Margutti}, {Levan}, {Perley}, {Fong}, {Wiersema}, {Menten}, \&
  {Hrudkova}}]{edo14}
{Laskar}, T., {Berger}, E., {Tanvir}, N., {Zauderer}, B.~A., {Margutti}, R.,
  {Levan}, A., {Perley}, D., {Fong}, W.-f., {Wiersema}, K., {Menten}, K., \&
  {Hrudkova}, M. 2014, \apj, 781, 1

\bibitem[{{Latif} {et~al.}(2013){Latif}, {Schleicher}, {Schmidt}, \&
  {Niemeyer}}]{latif13c}
{Latif}, M.~A., {Schleicher}, D.~R.~G., {Schmidt}, W., \& {Niemeyer}, J. 2013,
  \mnras, 433, 1607

\bibitem[{{MacFadyen} {et~al.}(2001){MacFadyen}, {Woosley}, \& {Heger}}]{mcf01}
{MacFadyen}, A.~I., {Woosley}, S.~E., \& {Heger}, A. 2001, \apj, 550, 410

\bibitem[{{Meiksin} \& {Whalen}(2013)}]{mw12}
{Meiksin}, A. \& {Whalen}, D.~J. 2013, \mnras, 430, 2854

\bibitem[{{Mesler} {et~al.}(2012){Mesler}, {Whalen}, {Lloyd-Ronning}, {Fryer},
  \& {Pihlstr{\"o}m}}]{met12a}
{Mesler}, R.~A., {Whalen}, D.~J., {Lloyd-Ronning}, N.~M., {Fryer}, C.~L., \&
  {Pihlstr{\"o}m}, Y.~M. 2012, \apj, 757, 117

\bibitem[{{M{\'e}sz{\'a}ros} \& {Rees}(2010)}]{mr10}
{M{\'e}sz{\'a}ros}, P. \& {Rees}, M.~J. 2010, \apj, 715, 967

\bibitem[{{Moderski} {et~al.}(2000){Moderski}, {Sikora}, \&
  {Bulik}}]{moderskiEA00}
{Moderski}, R., {Sikora}, M., \& {Bulik}, T. 2000, \apj, 529, 151

\bibitem[{{Moriya} {et~al.}(2013){Moriya}, {Blinnikov}, {Tominaga}, {Yoshida},
  {Tanaka}, {Maeda}, \& {Nomoto}}]{moriya12}
{Moriya}, T.~J., {Blinnikov}, S.~I., {Tominaga}, N., {Yoshida}, N., {Tanaka},
  M., {Maeda}, K., \& {Nomoto}, K. 2013, \mnras, 428, 1020

\bibitem[{{Nagakura} {et~al.}(2012){Nagakura}, {Suwa}, \& {Ioka}}]{nsi12}
{Nagakura}, H., {Suwa}, Y., \& {Ioka}, K. 2012, \apj, 754, 85

\bibitem[{{Nakar} \& {Granot}(2007)}]{ng07}
{Nakar}, E. \& {Granot}, J. 2007, \mnras, 380, 1744

\bibitem[{{Nakauchi} {et~al.}(2012){Nakauchi}, {Suwa}, {Sakamoto}, {Kashiyama},
  \& {Nakamura}}]{nak12}
{Nakauchi}, D., {Suwa}, Y., {Sakamoto}, T., {Kashiyama}, K., \& {Nakamura}, T.
  2012, \apj, 759, 128

\bibitem[{{Pan} {et~al.}(2012){Pan}, {Kasen}, \& {Loeb}}]{pan12a}
{Pan}, T., {Kasen}, D., \& {Loeb}, A. 2012, \mnras, 422, 2701

\bibitem[{Panaitescu \& Kumar(2000)}]{panaitescuKumar00}
Panaitescu, A. \& Kumar, P. 2000, \apj, 543, 66

\bibitem[{Panaitescu \& Meszaros(2000)}]{panaitescuMeszaros00}
Panaitescu, A. \& Meszaros, P. 2000, Astrophysical Journal Letters, 544

\bibitem[{{Passy} {et~al.}(2012){Passy}, {De Marco}, {Fryer}, {Herwig},
  {Diehl}, {Oishi}, {Mac Low}, {Bryan}, \& {Rockefeller}}]{pas12}
{Passy}, J.-C., {De Marco}, O., {Fryer}, C.~L., {Herwig}, F., {Diehl}, S.,
  {Oishi}, J.~S., {Mac Low}, M.-M., {Bryan}, G.~L., \& {Rockefeller}, G. 2012,
  \apj, 744, 52

\bibitem[{{Paul} {et~al.}(2011){Paul}, {Wei}, {Basa}, \& {Zhang}}]{paulEA11}
{Paul}, J., {Wei}, J., {Basa}, S., \& {Zhang}, S.-N. 2011, Comptes Rendus
  Physique, 12, 298

\bibitem[{{Pawlik} {et~al.}(2013){Pawlik}, {Milosavljevi{\'c}}, \&
  {Bromm}}]{pmb12}
{Pawlik}, A.~H., {Milosavljevi{\'c}}, M., \& {Bromm}, V. 2013, \apj, 767, 59

\bibitem[{{Pe'er}(2012)}]{peer12}
{Pe'er}, A. 2012, \apjl, 752, L8

\bibitem[{{Ritter} {et~al.}(2012){Ritter}, {Safranek-Shrader}, {Gnat},
  {Milosavljevi{\'c}}, \& {Bromm}}]{ritt12}
{Ritter}, J.~S., {Safranek-Shrader}, C., {Gnat}, O., {Milosavljevi{\'c}}, M.,
  \& {Bromm}, V. 2012, \apj, 761, 56

\bibitem[{{Roming} {et~al.}(2012){Roming}, {Pritchard}, {Prieto}, {Kochanek},
  {Fryer}, {Davidson}, {Humphreys}, {Bayless}, {Beacom}, {Brown}, {Holland},
  {Immler}, {Kuin}, {Oates}, {Pogge}, {Pojmanski}, {Stoll}, {Shappee},
  {Stanek}, \& {Szczygiel}}]{rom12}
{Roming}, P.~W.~A., {Pritchard}, T.~A., {Prieto}, J.~L., {Kochanek}, C.~S.,
  {Fryer}, C.~L., {Davidson}, K., {Humphreys}, R.~M., {Bayless}, A.~J.,
  {Beacom}, J.~F., {Brown}, P.~J., {Holland}, S.~T., {Immler}, S., {Kuin},
  N.~P.~M., {Oates}, S.~R., {Pogge}, R.~W., {Pojmanski}, G., {Stoll}, R.,
  {Shappee}, B.~J., {Stanek}, K.~Z., \& {Szczygiel}, D.~M. 2012, \apj, 751, 92

\bibitem[{Rybicki \& Lightman(1979)}]{rybickiLightman79}
Rybicki, G. \& Lightman, A. 1979, Radiative Processes in Astrophysics (New
  York: Wiley-Interscience)

\bibitem[{{Safranek-Shrader} {et~al.}(2013){Safranek-Shrader}, {Milosavljevic},
  \& {Bromm}}]{ss13}
{Safranek-Shrader}, C., {Milosavljevic}, M., \& {Bromm}, V. 2013,
  arXiv:1307.1982

\bibitem[{{Salvaterra} {et~al.}(2009){Salvaterra}, {Della Valle}, {Campana},
  {Chincarini}, {Covino}, {D'Avanzo}, {Fern{\'a}ndez-Soto}, {Guidorzi},
  {Mannucci}, {Margutti}, {Th{\"o}ne}, {Antonelli}, {Barthelmy}, {de Pasquale},
  {D'Elia}, {Fiore}, {Fugazza}, {Hunt}, {Maiorano}, {Marinoni}, {Marshall},
  {Molinari}, {Nousek}, {Pian}, {Racusin}, {Stella}, {Amati}, {Andreuzzi},
  {Cusumano}, {Fenimore}, {Ferrero}, {Giommi}, {Guetta}, {Holland}, {Hurley},
  {Israel}, {Mao}, {Markwardt}, {Masetti}, {Pagani}, {Palazzi}, {Palmer},
  {Piranomonte}, {Tagliaferri}, \& {Testa}}]{salv09}
{Salvaterra}, R., {Della Valle}, M., {Campana}, S., {Chincarini}, G., {Covino},
  S., {D'Avanzo}, P., {Fern{\'a}ndez-Soto}, A., {Guidorzi}, C., {Mannucci}, F.,
  {Margutti}, R., {Th{\"o}ne}, C.~C., {Antonelli}, L.~A., {Barthelmy}, S.~D.,
  {de Pasquale}, M., {D'Elia}, V., {Fiore}, F., {Fugazza}, D., {Hunt}, L.~K.,
  {Maiorano}, E., {Marinoni}, S., {Marshall}, F.~E., {Molinari}, E., {Nousek},
  J., {Pian}, E., {Racusin}, J.~L., {Stella}, L., {Amati}, L., {Andreuzzi}, G.,
  {Cusumano}, G., {Fenimore}, E.~E., {Ferrero}, P., {Giommi}, P., {Guetta}, D.,
  {Holland}, S.~T., {Hurley}, K., {Israel}, G.~L., {Mao}, J., {Markwardt},
  C.~B., {Masetti}, N., {Pagani}, C., {Palazzi}, E., {Palmer}, D.~M.,
  {Piranomonte}, S., {Tagliaferri}, G., \& {Testa}, V. 2009, \nat, 461, 1258

\bibitem[{Sari {et~al.}(1998)Sari, Piran, \& R.}]{sariEA98}
Sari, R., Piran, T., \& R., N. 1998, \apj, 497, L17

\bibitem[{{Smidt} {et~al.}(2014){Smidt}, {Whalen}, {Even}, {Wiggins},
  {Johnson}, \& {Fryer}}]{smidt13a}
{Smidt}, J., {Whalen}, D.~J., {Even}, W., {Wiggins}, B., {Johnson}, J.~L., \&
  {Fryer}, C.~L. 2014, arXiv:1401.5837

\bibitem[{{Smith} \& {Sigurdsson}(2007)}]{ss07}
{Smith}, B.~D. \& {Sigurdsson}, S. 2007, \apjl, 661, L5

\bibitem[{{Smith} {et~al.}(2009){Smith}, {Turk}, {Sigurdsson}, {O'Shea}, \&
  {Norman}}]{bsmith09}
{Smith}, B.~D., {Turk}, M.~J., {Sigurdsson}, S., {O'Shea}, B.~W., \& {Norman},
  M.~L. 2009, \apj, 691, 441

\bibitem[{{Spergel} {et~al.}(2013){Spergel}, {Gehrels}, {Breckinridge},
  {Donahue}, {Dressler}, {Gaudi}, {Greene}, {Guyon}, {Hirata}, {Kalirai},
  {Kasdin}, {Moos}, {Perlmutter}, {Postman}, {Rauscher}, {Rhodes}, {Wang},
  {Weinberg}, {Centrella}, {Traub}, {Baltay}, {Colbert}, {Bennett},
  {Kiessling}, {Macintosh}, {Merten}, {Mortonson}, {Penny}, {Rozo},
  {Savransky}, {Stapelfeldt}, {Zu}, {Baker}, {Cheng}, {Content}, {Dooley},
  {Foote}, {Goullioud}, {Grady}, {Jackson}, {Kruk}, {Levine}, {Melton},
  {Peddie}, {Ruffa}, \& {Shaklan}}]{spergelEA13}
{Spergel}, D., {Gehrels}, N., {Breckinridge}, J., {Donahue}, M., {Dressler},
  A., {Gaudi}, B.~S., {Greene}, T., {Guyon}, O., {Hirata}, C., {Kalirai}, J.,
  {Kasdin}, N.~J., {Moos}, W., {Perlmutter}, S., {Postman}, M., {Rauscher}, B.,
  {Rhodes}, J., {Wang}, Y., {Weinberg}, D., {Centrella}, J., {Traub}, W.,
  {Baltay}, C., {Colbert}, J., {Bennett}, D., {Kiessling}, A., {Macintosh}, B.,
  {Merten}, J., {Mortonson}, M., {Penny}, M., {Rozo}, E., {Savransky}, D.,
  {Stapelfeldt}, K., {Zu}, Y., {Baker}, C., {Cheng}, E., {Content}, D.,
  {Dooley}, J., {Foote}, M., {Goullioud}, R., {Grady}, K., {Jackson}, C.,
  {Kruk}, J., {Levine}, M., {Melton}, M., {Peddie}, C., {Ruffa}, J., \&
  {Shaklan}, S. 2013, arXiv:1305.5425

\bibitem[{{Stacy} {et~al.}(2011){Stacy}, {Bromm}, \& {Loeb}}]{stacy11b}
{Stacy}, A., {Bromm}, V., \& {Loeb}, A. 2011, \mnras, 413, 543

\bibitem[{{Stacy} {et~al.}(2012){Stacy}, {Greif}, \& {Bromm}}]{stacy12}
{Stacy}, A., {Greif}, T.~H., \& {Bromm}, V. 2012, \mnras, 422, 290

\bibitem[{Stanek {et~al.}(2003)Stanek, Matheson, Garnavich, Martini, Berlind,
  Caldwell, Challis, Brown, Schild, Krisciunas, Calkins, Lee, Jansen,
  Windhorst, Echevarria, Eisenstein, Pindor, Harding, Holland, \&
  Bersier}]{stanekEA03}
Stanek, K.~Z., Matheson, T., Garnavich, P.~M., Martini, P., Berlind, P.,
  Caldwell, N., Challis, P., Brown, W.~R., Schild, R., Krisciunas, K., Calkins,
  M.~L., Lee, J.~C.~Hathi, N., Jansen, R.~A., Windhorst, R., Echevarria, L.,
  Eisenstein, D.~J., Pindor, B.~Olszewski, E.~W., Harding, P., Holland, S.~T.,
  \& Bersier, D. 2003, \apj, 1, L17

\bibitem[{{Suwa} \& {Ioka}(2011)}]{suwa11}
{Suwa}, Y. \& {Ioka}, K. 2011, \apj, 726, 107

\bibitem[{{Tanaka} {et~al.}(2012){Tanaka}, {Moriya}, {Yoshida}, \&
  {Nomoto}}]{tet12}
{Tanaka}, M., {Moriya}, T.~J., {Yoshida}, N., \& {Nomoto}, K. 2012, \mnras,
  422, 2675

\bibitem[{{Taylor}(1950)}]{taylor50}
{Taylor}, G. 1950, Royal Society of London Proceedings Series A, 201, 159

\bibitem[{{Taylor} {et~al.}(2012){Taylor}, {Ellingson}, {Kassim}, {Craig},
  {Dowell}, {Wolfe}, {Hartman}, {Bernardi}, {Clarke}, {Cohen}, {Dalal},
  {Erickson}, {Hicks}, {Greenhill}, {Jacoby}, {Lane}, {Lazio}, {Mitchell},
  {Navarro}, {Ord}, {Pihlstr{\"o}m}, {Polisensky}, {Ray}, {Rickard},
  {Schinzel}, {Schmitt}, {Sigman}, {Soriano}, {Stewart}, {Stovall}, {Tremblay},
  {Wang}, {Weiler}, {White}, \& {Wood}}]{taylorEA12}
{Taylor}, G.~B., {Ellingson}, S.~W., {Kassim}, N.~E., {Craig}, J., {Dowell},
  J., {Wolfe}, C.~N., {Hartman}, J., {Bernardi}, G., {Clarke}, T., {Cohen}, A.,
  {Dalal}, N.~P., {Erickson}, W.~C., {Hicks}, B., {Greenhill}, L.~J., {Jacoby},
  B., {Lane}, W., {Lazio}, J., {Mitchell}, D., {Navarro}, R., {Ord}, S.~M.,
  {Pihlstr{\"o}m}, Y., {Polisensky}, E., {Ray}, P.~S., {Rickard}, L.~J.,
  {Schinzel}, F.~K., {Schmitt}, H., {Sigman}, E., {Soriano}, M., {Stewart},
  K.~P., {Stovall}, K., {Tremblay}, S., {Wang}, D., {Weiler}, K.~W., {White},
  S., \& {Wood}, D.~L. 2012, Journal of Astronomical Instrumentation, 1, 50004

\bibitem[{{Toma} {et~al.}(2011){Toma}, {Sakamoto}, \&
  {M{\'e}sz{\'a}ros}}]{toma11}
{Toma}, K., {Sakamoto}, T., \& {M{\'e}sz{\'a}ros}, P. 2011, \apj, 731, 127

\bibitem[{{Tominaga} {et~al.}(2011){Tominaga}, {Morokuma}, {Blinnikov},
  {Baklanov}, {Sorokina}, \& {Nomoto}}]{tomin11}
{Tominaga}, N., {Morokuma}, T., {Blinnikov}, S.~I., {Baklanov}, P., {Sorokina},
  E.~I., \& {Nomoto}, K. 2011, \apjs, 193, 20

\bibitem[{{Totani}(1997)}]{tot97}
{Totani}, T. 1997, \apjl, 486, L71

\bibitem[{{Totani} {et~al.}(2006){Totani}, {Kawai}, {Kosugi}, {Aoki}, {Yamada},
  {Iye}, {Ohta}, \& {Hattori}}]{tot06}
{Totani}, T., {Kawai}, N., {Kosugi}, G., {Aoki}, K., {Yamada}, T., {Iye}, M.,
  {Ohta}, K., \& {Hattori}, T. 2006, \pasj, 58, 485

\bibitem[{{Turk} {et~al.}(2009){Turk}, {Abel}, \& {O'Shea}}]{turk09}
{Turk}, M.~J., {Abel}, T., \& {O'Shea}, B. 2009, Science, 325, 601

\bibitem[{{Wang} {et~al.}(2012){Wang}, {Bromm}, {Greif}, {Stacy}, {Dai},
  {Loeb}, \& {Cheng}}]{wang12}
{Wang}, F.~Y., {Bromm}, V., {Greif}, T.~H., {Stacy}, A., {Dai}, Z.~G., {Loeb},
  A., \& {Cheng}, K.~S. 2012, \apj, 760, 27

\bibitem[{{Wang} {et~al.}(2011){Wang}, {Qi}, \& {Dai}}]{wang11}
{Wang}, F.-Y., {Qi}, S., \& {Dai}, Z.-G. 2011, \mnras, 415, 3423

\bibitem[{{Whalen} {et~al.}(2004){Whalen}, {Abel}, \& {Norman}}]{wan04}
{Whalen}, D., {Abel}, T., \& {Norman}, M.~L. 2004, \apj, 610, 14

\bibitem[{{Whalen} \& {Norman}(2006)}]{wn06}
{Whalen}, D. \& {Norman}, M.~L. 2006, \apjs, 162, 281

\bibitem[{{Whalen} \& {Norman}(2008{\natexlab{a}})}]{wn08b}
---. 2008{\natexlab{a}}, \apj, 673, 664

\bibitem[{{Whalen} {et~al.}(2008{\natexlab{a}}){Whalen}, {Prochaska}, {Heger},
  \& {Tumlinson}}]{wet08c}
{Whalen}, D., {Prochaska}, J.~X., {Heger}, A., \& {Tumlinson}, J.
  2008{\natexlab{a}}, \apj, 682, 1114

\bibitem[{{Whalen} {et~al.}(2008{\natexlab{b}}){Whalen}, {van Veelen},
  {O'Shea}, \& {Norman}}]{wet08a}
{Whalen}, D., {van Veelen}, B., {O'Shea}, B.~W., \& {Norman}, M.~L.
  2008{\natexlab{b}}, \apj, 682, 49

\bibitem[{{Whalen}(2012)}]{dw12}
{Whalen}, D.~J. 2012, arXiv:1209.4688

\bibitem[{{Whalen} {et~al.}(2013{\natexlab{a}}){Whalen}, {Even}, {Frey},
  {Smidt}, {Johnson}, {Lovekin}, {Fryer}, {Stiavelli}, {Holz}, {Heger},
  {Woosley}, \& {Hungerford}}]{wet12b}
{Whalen}, D.~J., {Even}, W., {Frey}, L.~H., {Smidt}, J., {Johnson}, J.~L.,
  {Lovekin}, C.~C., {Fryer}, C.~L., {Stiavelli}, M., {Holz}, D.~E., {Heger},
  A., {Woosley}, S.~E., \& {Hungerford}, A.~L. 2013{\natexlab{a}}, \apj, 777,
  110

\bibitem[{{Whalen} {et~al.}(2013{\natexlab{b}}){Whalen}, {Even}, {Lovekin},
  {Fryer}, {Stiavelli}, {Roming}, {Cooke}, {Pritchard}, {Holz}, \&
  {Knight}}]{wet12e}
{Whalen}, D.~J., {Even}, W., {Lovekin}, C.~C., {Fryer}, C.~L., {Stiavelli}, M.,
  {Roming}, P.~W.~A., {Cooke}, J., {Pritchard}, T.~A., {Holz}, D.~E., \&
  {Knight}, C. 2013{\natexlab{b}}, \apj, 768, 195

\bibitem[{{Whalen} {et~al.}(2013{\natexlab{c}}){Whalen}, {Even}, {Smidt},
  {Heger}, {Chen}, {Fryer}, {Stiavelli}, {Xu}, \& {Joggerst}}]{wet12d}
{Whalen}, D.~J., {Even}, W., {Smidt}, J., {Heger}, A., {Chen}, K.-J., {Fryer},
  C.~L., {Stiavelli}, M., {Xu}, H., \& {Joggerst}, C.~C. 2013{\natexlab{c}},
  \apj, 778, 17

\bibitem[{{Whalen} {et~al.}(2013{\natexlab{d}}){Whalen}, {Even}, {Smidt},
  {Heger}, {Hirschi}, {Yusof}, {Stiavelli}, {Fryer}, {Chen}, \&
  {Joggerst}}]{wet13e}
{Whalen}, D.~J., {Even}, W., {Smidt}, J., {Heger}, A., {Hirschi}, R., {Yusof},
  N., {Stiavelli}, M., {Fryer}, C.~L., {Chen}, K.-J., \& {Joggerst}, C.~C.
  2013{\natexlab{d}}, arXiv:1312.5360

\bibitem[{{Whalen} \& {Fryer}(2012)}]{wf12}
{Whalen}, D.~J. \& {Fryer}, C.~L. 2012, \apjl, 756, L19

\bibitem[{{Whalen} {et~al.}(2013{\natexlab{e}}){Whalen}, {Fryer}, {Holz},
  {Heger}, {Woosley}, {Stiavelli}, {Even}, \& {Frey}}]{wet12a}
{Whalen}, D.~J., {Fryer}, C.~L., {Holz}, D.~E., {Heger}, A., {Woosley}, S.~E.,
  {Stiavelli}, M., {Even}, W., \& {Frey}, L.~H. 2013{\natexlab{e}}, \apjl, 762,
  L6

\bibitem[{{Whalen} {et~al.}(2013{\natexlab{f}}){Whalen}, {Joggerst}, {Fryer},
  {Stiavelli}, {Heger}, \& {Holz}}]{wet12c}
{Whalen}, D.~J., {Joggerst}, C.~C., {Fryer}, C.~L., {Stiavelli}, M., {Heger},
  A., \& {Holz}, D.~E. 2013{\natexlab{f}}, \apj, 768, 95

\bibitem[{{Whalen} {et~al.}(2013{\natexlab{g}}){Whalen}, {Johnson}, {Smidt},
  {Heger}, {Even}, \& {Fryer}}]{wet13b}
{Whalen}, D.~J., {Johnson}, J.~L., {Smidt}, J., {Heger}, A., {Even}, W., \&
  {Fryer}, C.~L. 2013{\natexlab{g}}, \apj, 777, 99

\bibitem[{{Whalen} {et~al.}(2013{\natexlab{h}}){Whalen}, {Johnson}, {Smidt},
  {Meiksin}, {Heger}, {Even}, \& {Fryer}}]{wet13a}
{Whalen}, D.~J., {Johnson}, J.~L., {Smidt}, J., {Meiksin}, A., {Heger}, A.,
  {Even}, W., \& {Fryer}, C.~L. 2013{\natexlab{h}}, \apj, 774, 64

\bibitem[{{Whalen} \& {Norman}(2008{\natexlab{b}})}]{wn08a}
{Whalen}, D.~J. \& {Norman}, M.~L. 2008{\natexlab{b}}, \apj, 672, 287

\bibitem[{{Whalen} {et~al.}(2014){Whalen}, {Smidt}, {Even}, {Woosley}, {Heger},
  {Stiavelli}, \& {Fryer}}]{wet13d}
{Whalen}, D.~J., {Smidt}, J., {Even}, W., {Woosley}, S.~E., {Heger}, A.,
  {Stiavelli}, M., \& {Fryer}, C.~L. 2014, \apj, 781, 106

\bibitem[{{Whalen} {et~al.}(2013{\natexlab{i}}){Whalen}, {Smidt}, {Johnson},
  {Holz}, {Stiavelli}, \& {Fryer}}]{wet13c}
{Whalen}, D.~J., {Smidt}, J., {Johnson}, J.~L., {Holz}, D.~E., {Stiavelli}, M.,
  \& {Fryer}, C.~L. 2013{\natexlab{i}}, arXiv:1312.6330

\bibitem[{Wijers \& Galama(1999)}]{wijersGalama99}
Wijers, R. \& Galama, T. 1999, \apj, 523, 177

\bibitem[{{Wijers} {et~al.}(1998){Wijers}, {Bloom}, {Bagla}, \&
  {Natarajan}}]{wij98}
{Wijers}, R.~A.~M.~J., {Bloom}, J.~S., {Bagla}, J.~S., \& {Natarajan}, P. 1998,
  \mnras, 294, L13

\bibitem[{{Wise} {et~al.}(2012){Wise}, {Turk}, {Norman}, \& {Abel}}]{wise12}
{Wise}, J.~H., {Turk}, M.~J., {Norman}, M.~L., \& {Abel}, T. 2012, \apj, 745,
  50

\bibitem[{{Woosley}(1993)}]{woo93}
{Woosley}, S.~E. 1993, \apj, 405, 273

\bibitem[{{Woosley} \& {Bloom}(2006)}]{woo06}
{Woosley}, S.~E. \& {Bloom}, J.~S. 2006, \araa, 44, 507

\bibitem[{{Yoon} \& {Langer}(2005)}]{yoon05}
{Yoon}, S.-C. \& {Langer}, N. 2005, \aap, 443, 643

\bibitem[{{Yoon} {et~al.}(2006){Yoon}, {Langer}, \& {Norman}}]{yoon06}
{Yoon}, S.-C., {Langer}, N., \& {Norman}, C. 2006, \aap, 460, 199

\bibitem[{{Zhang} \& {Fryer}(2001)}]{zf01}
{Zhang}, W. \& {Fryer}, C.~L. 2001, \apj, 550, 357

\end{thebibliography}

\appendix

In the canonical fireball model, GRBs are highly relativistic jets that propagate 
adiabatically into an ambient medium, i.e., only a very small fraction of the total 
energy of the burst is radiated away by electrons.  Our afterglow model is valid 
in both relativistic and non-relativistic regimes and can model emission from 
jets in a variety of circumstellar media, including dense shells. It is an extension 
to the method in M12, and now incorporates radiative blast waves, spherical 
emission and beaming, and inverse Compton scattering (but does not include 
pair production).  Our method is an important improvement over past work in 
that we can now smoothly evolve energy conservation for the jet from its usual 
form (Equation (1) below) from the highly relativistic regime to mildly relativistic 
and Newtonian regimes (Equations (11) - (14) below) as the jet decelerates 
through shells and other abrupt dense obstacles in its path.  Our method 
produces K band light curves for Pop III GRBs in uniform \HII\ regions at $z \sim
$ 20 that are in excellent agreement with those of \citet{gou04}.  In what follows, 
primed quantities refer to the frame of the jet, unprimed quantities refer to the 
frame of the surrounding interstellar medium, and the subscript $\earth$ refers 
to the Earth frame.

The GRB blast wave is modeled as a uniform jet with an initial half-opening 
angle $\theta_0$ and Lorentz factor $\Gamma_0$.  The kinetic energy of the 
jet can be determined from $E_{\mathrm{iso},\gamma}$, $\Gamma_0$, and 
$\theta_0$.  Energy conservation yields the Lorentz factor of the jet, $\Gamma$, 
as it propagates through the external medium:

\begin{equation}
\frac{d\Gamma}{dm} = -\frac{\hat{\gamma}\left(\Gamma^2 - 1\right) - \left(\hat{
\gamma} - 1\right)\Gamma\beta^2}{M_\text{ej} + \epsilon m + (1-\epsilon)m\left[
2\hat{\gamma}\Gamma - \left(\hat{\gamma} - 1\right)\left(1 + \Gamma^{-2}\right)
\right]},
\label{eqn:GammaEquation}
\end{equation}

\noindent where $M_\text{ej}$ is the initial mass of the jet ejecta, $m$ is the total 
mass that has been swept up by the jet, $\beta = \left(1-\Gamma^{-2}\right)^{1/2}
$ in the normalized bulk velocity, $\hat{\gamma} \simeq (4\Gamma+1)/(3\Gamma)
$ is the adiabatic index \citep{huangEA00}, and $\epsilon$ is the radiative efficiency 
\citep{peer12}.  $\epsilon$ in turn is given by \citep{daiEA99}

\begin{equation}
\epsilon = \epsilon_e\frac{{t'}_\text{syn}^{-1}}{{t'}_\text{syn}^{-1} + {t'}_\text{exp}^{-1}}, 
\end{equation}

\noindent where $\epsilon_e$ is the fraction of the burst energy stored in the 
electrons, $t'_\text{syn}$ is the synchrotron cooling time scale of the injected 
electrons, and $t'_\text{exp}$ is the age of the remnant. 

The high resolution of our profiles allows us to treat the density as constant across 
each grid point.  In the canonical GRB model, the jet is also assumed to sweep up 
all the material in its path.  The total mass swept up by the jet by the time it reaches 
grid point $n$ is therefore approximately

\begin{equation}
m(r) = \frac{4}{3}\pi\left(\rho_1 r_1^3 + \sum_{i=2}^{n}{\rho_i\left(r_i^3 - r_{i-1}^3\right)} 
\right),
\end{equation}
\noindent where $r_i$ and $\rho_i$ are the radius and density of the $i$th grid 
point, respectively.  The time $t_\text{obs}$ at which a photon emitted at the 
leading edge of the jet reaches an observer along the line of sight can be found 
by integrating Equation (12) from \citet{huangEA99}:

\begin{equation}
t = \frac{1}{c}\int{\frac{dr}{\beta\Gamma\left(\Gamma + \sqrt{\Gamma^2 - 1}\right),}}\, 
\label{eqn:tObsEquation}
\end{equation}

\noindent where $c$ is the speed of light.

As the jet propagates toward the observer, it also expands laterally at the 
comoving sound speed

\begin{equation}
c'_s = \frac{da'}{dt'} = \sqrt{\frac{\hat{\gamma}\left(\hat{\gamma}-1\right)\left(\Gamma
-1\right)}{1+\hat{\gamma}\left(\Gamma-1\right)}},
\end{equation}

\noindent where $a$ is the half the diameter of the leading edge of the jet.  
Transforming into the isotropic frame, we get

\begin{equation}
\frac{da'}{dt'} = \frac{1}{\Gamma}\frac{da}{dt}.
\label{eqn:dAdTequation}
\end{equation}

Because the jet half-opening angle $\theta_j = \arctan(a/r)$, the factor of $1/
\Gamma$ in Equation (\ref{eqn:dAdTequation}) leads to a nearly constant 
value of $\theta_j$ until the Lorentz factor approaches $\Gamma = 1/\theta
_j$, when the jet experiences rapid lateral expansion in the isotropic frame.  
It is somewhat standard practice to use the approximation $\theta_j \simeq 
a/r$, but this expression is not valid at late times, when $\theta_j$ can be 
greater than $30^\circ$.  

At the onset of the afterglow, the jet can be modeled as a slab of thickness 
$\Delta = ct_\text{B}$, where $c$ is the speed of light and $t_\text{B}$ is the 
isotropic frame burst duration.  The thickness of the jet in the isotropic frame 
is related to its thickness in the comoving frame by $\Delta = \Delta'/\Gamma
$.  For a differential comoving frame time $\delta t'$, the evolution of the jet 
thickness is $\delta\Delta = c'_s\delta t'/\Gamma$.  The relationship between 
$\delta t'$ and $\delta t$ is $\delta t = \Gamma\delta t'$.  Therefore, the 
isotropic frame jet thickness evolves as

\begin{equation}
\frac{d\Delta}{dt} = \frac{c'_s}{\Gamma^2}
\label{eqn:jetThickness}
\vspace{0.1in}
\end{equation}

\subsection{Density Jumps}

If the GRB jet encounters an abrupt change in the density of the surrounding
medium, Equation (\ref{eqn:GammaEquation}), which assumes self-similar 
expansion of the ejecta, no longer applies.  When the jet collides with the 
jump, a contact discontinuity forms between the material that was swept up 
by the jet prior to the collision and the material that was swept up afterwards.  
A reverse shock forms at the contact discontinuity and backsteps into the jet
in the frame of the jet.  A forward shock also forms just past the jump and 
propagates into the medium beyond it. The contact discontinuity also moves 
forward, but at a much lower velocity as the jet continues to advance into the 
new medium and plow it up.

Although several analytic treatments exist for the evolution of the jet and its 
component shocks at a density jump \citep{ng07, daiLu02, gatEA13}, they all 
assume that the jet will be relativistic upon collision with the jump, which is not 
always true for our density profiles. We must therefore resort to the general 
\citet{blandfordMcKee76} and \citet{taylor50} equations for the jump conditions 
in order to self-consistently evolve the jet evolution in any medium.

\begin{figure}
\plotone{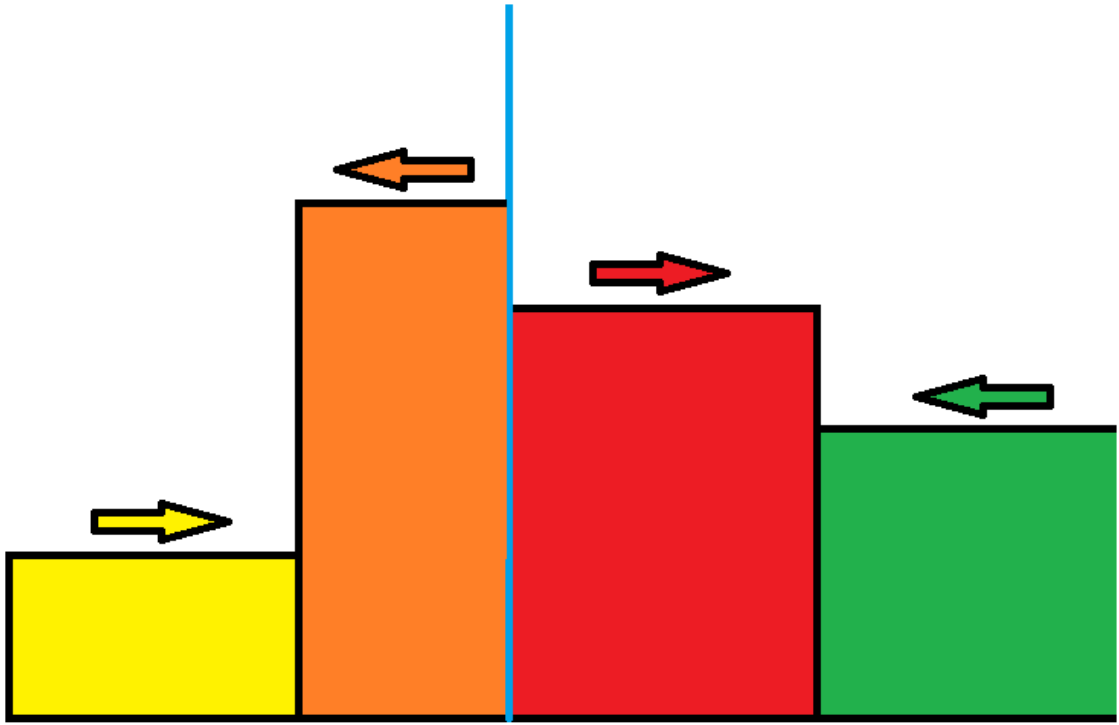} 
\caption{The interaction of a GRB jet with a density jump.  The medium on 
either side of the jump can be broken into four regions: region I (green), the 
undisturbed material beyond the density jump, region II (red), the material 
that has been shocked by the forward shock, region III (orange), the region 
of the jet through which the reverse shock has passed, and region IV (yellow), 
the unshocked jet material.  A contact discontinuity forms between regions II 
and III (blue line), effectively preventing the material in those two regions from 
mixing.  The colored arrows show the direction of flow of the four regions in 
the frame of the contact discontinuity.  The height of each colored block 
represents the density of that region (in arbitrary units).}
\vspace{0.1in}
\label{fig:densityJump}
\end{figure}

When the jet encounters a density jump, the medium surrounding the jump 
can be partitioned into four regions (Figure~\ref{fig:densityJump}).  Region 
I is the undisturbed material beyond the density jump.  Region II is the 
postshock material behind the forward shock, region III is the region of the 
jet through which the reverse shock has passed, and region IV is the 
unshocked jet material.  The contact discontinuity forms between regions II 
and III.  For a relativistic or mildly-relativistic jet, the jump conditions follow 
from the conservation of energy ($w\gamma^2\beta$), momentum ($w
\gamma^2\beta^2 + p$), and particle number ($n\gamma\beta$) flux 
densities across the jump in the frame of the shock, where $w$ and $p$ are 
the enthalpy and pressure of the gas, respectively, and $\gamma$ and 
$\beta$ are the Lorentz factor and velocity of the gas particles. The pressure 
of the gas is 

\begin{equation}
p = \left(\hat{\gamma} - 1\right)\left(e - \rho\right),
\label{eqn:condition1}
\end{equation}

\noindent where $\hat{\gamma}$ is the adiabatic index,$e = \gamma nm_pc
^2$ is the energy density of the gas and $\rho = nm_pc^2$ is the rest-frame 
gas density.  The enthalpy $w = e + p$.     

For any two adjacent regions 1 and 2 in regions I-IV described above, the 
jump conditions for relativistic and mildly-relativistic jets are 
\citep{blandfordMcKee76}

\begin{equation}
\frac{e_2}{n_2} = \gamma_2\frac{w_1}{n_1},
\label{eqn:condition2}
\end{equation}

\begin{equation}
\frac{n_2}{n_1} = \frac{\hat{\gamma_2}\gamma_2 + 1}{\hat{\gamma_2} - 1},
\label{eqn:condition3}
\end{equation}

\noindent and

\begin{equation}
\Gamma^2 = \frac{\left(\gamma_2 + 1\right)\left[\hat{\gamma_2}\left(\gamma_2 
- 1\right) + 1\right]^2}{\hat{\gamma_2}\left(2 - \hat{\gamma_2}\right)\left(
\gamma_2 - 1\right) + 2},
\end{equation}

\noindent where $\Gamma$ is the Lorentz factor of the new forward shock.  
There are three known quantities: the Lorentz factor of the jet when it crashes 
into the density jump and the densities $n_I$ and $n_{IV}$ of regions I and IV, 
respectively.  From these three quantities and the requirement that the pressure 
and energy density must be equal in regions II and III (due to the presence of a 
contact discontinuity between them) the Lorentz factors of the forward shock, the 
reverse shock, and the contact discontinuity can be determined from the three 
jump conditions.

Given these Lorentz factors an infinitesimal time after the leading edge of the
jet collides with the jump, the subsequent motion of the forward shock is 
determined from energy conservation with Equation (4) of \citet{ng07}.  The 
evolution of the reverse shock is set by the fact that the energy density remains 
constant across the contact discontinuity.  The Lorentz factor that satisfies this
requirement is found numerically.  Meanwhile, the initial shock, which has no
knowledge of the contact discontinuity, continues to expand adiabatically.

Previous authors have assumed the jet to be relativistic when it encounters the 
jump.  This in general is true for simple density profiles where one or perhaps a 
few density jumps are present.  In our much more complicated and realistic 
profiles, however, there can be dozens of density jumps. If the jump scale factor 
$a = n/n_0$ (where $n_0$ and $n$ are the densities of the medium before and 
after the jump, respectively) is $>$ 1, the Lorentz factor of the new forward 
shock will usually be lower than that of the initial shock. If the new forward shock 
encounters a second density jump, then another forward shock will be produced 
with an even lower Lorentz factor.  The forward shock can eventually become 
non-relativistic when it encounters a subsequent jump, and the Blandford-McKee 
conditions will no longer apply.  The flow will instead obey the Sedov-von 
Neumann-Taylor jump conditions for a Newtonian fluid \citep{taylor50}:

\begin{equation}
 \frac{\rho_1}{\rho_0} = \frac{\hat{\gamma} - 1 + \left(\hat{\gamma} + 1\right)y_1}{
 \hat{\gamma} + 1 + \left(\hat{\gamma} - 1\right)y_1},
 \label{eqn:NRcondition1}
\end{equation}

\begin{equation}
 \frac{U^2}{a^2} = \frac{1}{2\hat{\gamma}}\left[\hat{\gamma} - 1 + 
 \left(\hat{\gamma} + 1\right)y_1\right],
 \label{eqn:NRcondition2}
\end{equation}

\noindent and

\begin{equation}
 \frac{u_1}{U} = \frac{2\left(y_1 - 1\right)}{\hat{\gamma} - 1 + \left(\hat{\gamma} 
 + 1\right)y_1},
 \label{eqn:NRcondition3}
\end{equation}

\noindent where $\rho_0$ and $\rho_1$ are the densities before and after the 
jump, respectively, $\hat{\gamma} = 5/3$ is the adiabatic index, $y_1 = p_1/p_0
$, $p_0$ and $p_1$ are the pressures behind and ahead of the jump, $a$ is the 
sound speed, $U$ is the velocity of the forward shock, and $u_1$ is the velocity 
of the fluid behind the forward shock.  The velocities of the forward and reverse 
shocks and the contact discontinuity can be found by solving these jump 
conditions numerically if the velocity of the initial shock and the pressures of 
regions I and IV are known.  Our ZEUS-MP models provide the temperature of 
region IV at each grid point, and the temperature of region I can be calculated 
by assuming equipartition of thermal and kinetic energies within the jet.  The
pressures in these regions are then easily obtained from the temperatures by
the ideal gas law.

Our afterglow model in M12 essentially ignores the presence of density jumps
and instead assumes that the evolution of the jet is dependent only on the total
mass it sweeps up.  The mass swept up by the jet is assumed to instantly mix 
with previously swept-up material, with a prompt change in the Lorentz factor 
along the entire jet.  In reality, any collision with external structures produces a 
reverse shock that steps backward through the jet.  Until the reverse shock 
reaches a location in the jet, that location is not aware that any interaction has 
taken place.  Consequently, any sharp feature in the light curve that might be 
produced by a density jump will be somewhat smoothed out by a time equal to 
the jet-crossing time of the reverse shock.

\subsection{The Injection Break}

If we assume that a constant fraction $\epsilon_B$ of the total fireball energy is 
stored in magnetic fields, then the equipartition magnetic field strength at the 
shock boundary is \citep[i.e.][]{huangEA00}:

\begin{equation}
\frac{B'^2}{8\pi} = \epsilon_B \frac{\hat{\gamma}\Gamma + 1}{\hat{\gamma} - 1}
\left(\Gamma - 1\right)n(r)m_pc^2,
\end{equation}   

\noindent where $\epsilon_B$ is the fraction of the burst energy stored in magnetic 
fields, $n(r)$ is the number density of the medium at radius $r$, and $m_p$ is the 
proton 
mass.

The electrons that are injected into the shock are assumed to have a velocity 
distribution $N(\gamma) \propto \gamma^{-p}$ with a minimum Lorentz factor 
$\gamma_m$.  The minimum Lorentz factor of the injected electrons is \citep{
huangEA00}

\begin{equation}
\gamma_m = \epsilon_e\left(\Gamma - 1\right)\frac{m_p\left(p-2\right)}{m_e
\left(p-1\right)} + 1,
\end{equation}

\noindent where $m_e$ is the electron mass.  Electrons with a Lorentz factor 
$\gamma_e$ emit synchrotron radiation at a characteristic frequency \citep{
rybickiLightman79}

\begin{equation}
\nu(\gamma_e) = \frac{\gamma}{1 - \beta_e} \frac{3q_eB}{4\pi m_ec} 
\label{eqn:nu_e},
\end{equation}

\noindent where $\beta_e = \sqrt{1 - \gamma_e^{-2}}$, and $q_e$ is the electron 
charge.  The injection break frequency $\nu_m$ corresponds to the peak emission 
frequency of the injected electrons, and can be found by substituting $\gamma_e 
= \gamma_m$ into Equation (\ref{eqn:nu_e}).

\subsection{The Cooling Break}

Relativistic electrons in the shock cool radiatively through inverse Compton (IC) 
scattering and synchrotron emission on a comoving frame timescale \citep{
panaitescuKumar00}

\begin{equation}
t'_\text{rad}(\gamma) = \frac{t'_\text{syn}}{Y+1}, \label{eqn:remnantAgeEquation}
\end{equation}

\noindent where $Y$ is the Compton parameter.  The synchrotron cooling time 
scale $t'_\text{syn}$ of an electron is equal to the ratio of its energy $E$ to the 
synchrotron power $P$ it radiates:

\begin{equation}
t'_\text{syn} = \frac{E}{P} = \frac{1}{\gamma_e\beta_e^2} \frac{6\pi m_e c}{\sigma
_T B'^2},
\label{eqn:t_syn}
\end{equation}

\noindent where $\sigma_T$ is the Thomson cross section for electron scattering.
Equations (\ref{eqn:remnantAgeEquation}) and (\ref{eqn:t_syn}) can be solved to
find the Lorentz factor for electrons that cool on a timescale equal to the observer
frame age of the remnant, $\gamma_c$:

\begin{equation}
\gamma_c = \frac{6\pi m_ec^2}{B'^2\sigma_e(Y + 1)t'}. 
\label{eqn:coolingBreakEquation}
\end{equation}

The frequency of the cooling break, $\nu_c$, is found by substituting $\gamma_e
= \gamma_c$ in Equation (\ref{eqn:nu_e}).

\subsubsection{Fast-Cooling Electrons}

Electrons in the jet can also cool by adiabatic expansion of the gas. When cooling 
times for electrons with Lorentz factor $\gamma_m$ are less than the age of the 
jet ($\nu_c < \nu_m$, where $\nu_c$ is the frequency of the cooling break), the 
electrons in the jet lose a significant portion of their energy through emission of 
radiation and are said to be radiative, or fast-cooling.  On the other hand, if the 
cooling time is greater than the age of the jet ($\nu_c > \nu_m$) the electrons do 
not lose much energy to radiation and are said to be adiabatic, or slow-cooling.

To calculate $Y$, we only account for one upscattering of the synchrotron photons.  
If the electrons injected into the shock are fast-cooling and the frequency of the 
absorption break $\nu_a < \text{min}(\nu_m, \nu_c)$, then $Y$ is approximately 
\citep{panaitescuMeszaros00}

\begin{equation}
Y_r = \gamma_\text{m}\gamma_\text{c}\tau_\text{e}, 
\label{eqn:radiativeComptonParameterEquation1}
\end{equation}

\noindent where a constant of order unity is ignored and $\tau_e$, the optical 
depth to electron scattering, is 

\begin{equation}
\tau'_e = \frac{\sigma_em(r)}{4\pi m_\text{p}r'^2}.
\end{equation}

\noindent The medium becomes optically thick to synchrotron self-absorption at 
the absorption break frequency, $\nu_a$.  When both the injection break and the 
cooling break lie in the optically thick regime, $Y$ becomes

\begin{equation}
Y_r = Y_* = \tau'_e\left(C_2^{2-p}\gamma_c^7\gamma_m^{7(p-1)}\right)^{1/(p+5)},
\end{equation}

\noindent where $C_2 \equiv 5q_e\tau'_e/\sigma_eB'$ 
\citep{panaitescuMeszaros00}. 

\subsubsection{Slow-Cooling Electrons}

If the electrons are slow-cooling, $Y$ becomes

\begin{equation}
Y_a = \tau'_\text{e}\gamma_\text{i}^{p-1}\gamma_\text{c}^{3-p},
\end{equation}

\noindent as long as $\nu_a < \text{min}(\nu_m, \nu_c)$ \citep{
panaitescuMeszaros00} and $1 < p < 3$ (here, we have again ignored a constant 
of order unity).  If the injection and cooling breaks lie in the region of the spectrum 
that is optically thick to synchrotron self-absorption, then $Y$ is the same as in the 
corresponding fast-cooling case and

\begin{equation}
Y_a = Y_*.
\end{equation}

\subsection{The Absorption Break}

At lower frequencies, the medium in which the jet propagates becomes optically 
thick to synchrotron self-absorption.  $F_\nu$ then becomes $\propto \nu^2$ at 
some absorption break frequency $\nu_a$ where the optical depth to 
self-absorption is $\tau_\text{ab} = 1$.  The frequency of the absorption break 
depends on the cooling regime of the electrons (fast or slow) and on the order 
and values of the injection and cooling breaks.  In the fast-cooling regime \citep{
panaitescuMeszaros00}, 

\begin{equation}
\nu'_\text{a, fast-cooling} = 
\begin{cases}
C_2^{3/10}\gamma_c^{-1/2}       ,                 & \gamma_a < \gamma_c < \gamma_m \\
\left(C_2\gamma_c\right)^{1/6} ,                  & \gamma_c < \gamma_a < \gamma_m \\
\left(C_2\gamma_c\gamma_m^{p-1}\right)^{1/(p+5)}, & \gamma_c < \gamma_m < \gamma_a, \\
\end{cases}
\end{equation}

\noindent and in the slow-cooling regime

\begin{equation}
\nu'_\text{a, slow-cooling} = 
\begin{cases}
C_2^{3/10}\gamma_m^{-1/2}       ,                  & \gamma_a < \gamma_m < \gamma_c \\
\left(C_2\gamma_m^{p-1}\right)^{1/(p-4)} ,        & \gamma_m < \gamma_a < \gamma_c \\
\left(C_2\gamma_m^{p-1}\gamma_c\right)^{1/(p+5)}, & \gamma_m < \gamma_c < \gamma_a. \\
\end{cases}
\end{equation}

\subsection{Light Curves}

In order to produce light curves, we must first find the time dependence of 
$\Gamma(r)$, $n(r)$, and $m(r)$.  Equation (\ref{eqn:GammaEquation}) can be 
solved numerically for $\Gamma(r)$, and Equation (\ref{eqn:tObsEquation}) can
then be used to relate the observer time $t_\text{obs}$ to the jet position $r$, 
allowing us to rewrite the equations defining the three break frequencies in terms 
of $t_\text{obs}$, $\Gamma(t_\text{obs})$, $n(t_\text{obs})$, and $M(t_\text{obs})
$.  Given the three break frequencies and the peak flux density, analytical light 
curves can then be calculated from the radio to the $\gamma$-ray regions of the 
spectrum.  If $\nu_a < \text{min}(\nu_m, \nu_c)$, then the peak flux density $F_{
\nu,\earth}^\text{max}$ occurs at the injection break if $\nu_\text{m} < \nu_\text{c}
$ and at the cooling break if $\nu_\text{m} > \nu_\text{c}$:

\begin{equation}
F_{\nu,\earth}^\text{max} = \frac{\sqrt{3}\phi_\text{p}}{4\pi D^2}\frac{q_e^3\beta_
m^2}{m_\text{e}c^2}\frac{\Gamma B'm(r)}{m_\text{p}},
\end{equation}

\noindent where $\phi_\text{p}$ is a factor calculated by \citet{wijersGalama99} 
that depends on the value of p, and $D = (1+z)^{-1/2}D_l$, where $D_l$ is the 
luminosity distance to the source \citep{panaitescuKumar00}.  The flux at any 
frequency $\nu$ (ignoring relativistic beaming and the spherical nature of the 
emitting region) has been derived by \citet{sariEA98} and \citet{
panaitescuKumar00} as described below.  

\subsubsection{Fast-Cooling Electrons}

When the electrons are in the fast-cooling regime, the peak flux occurs at the 
cooling break as long as $\nu_a < \nu_c$:

\begin{equation}
F_{\nu, \earth} = F_{\nu,\earth}^\text{max}
\begin{cases}
(\nu/\nu_a)^2(\nu_a/\nu_c)^{1/3},       &         \nu < \nu_a \\
(\nu/\nu_c)^{1/3},                      & \nu_a < \nu < \nu_c \\
(\nu/\nu_c)^{-1/2},                     & \nu_c < \nu < \nu_m \\
(\nu/\nu_m)^{-p/2}(\nu_m/\nu_c)^{-1/2}, & \nu_m < \nu.
\end{cases}
\label{eqn:fluxEquation1}
\end{equation}

\noindent If the medium is optically thick to synchrotron self-absorption at the 
cooling break frequency, the maximum flux moves to the absorption break 
frequency.  Between the absorption and cooling breaks, $F_\nu \propto \nu^{
5/2}$ but becomes $\propto \nu^{2}$ below the cooling break:

\begin{equation}
F_{\nu, \earth} = F_{\nu,\earth}^\text{max}
\begin{cases}
(\nu/\nu_c)^2(\nu_c/\nu_a)^{5/2},       &         \nu < \nu_c \\
(\nu/\nu_a)^{5/2},                      & \nu_c < \nu < \nu_a \\
(\nu/\nu_a)^{-1/2},                     & \nu_a < \nu < \nu_m \\
(\nu/\nu_m)^{-p/2}(\nu_m/\nu_a)^{-1/2}, & \nu_m < \nu.
\end{cases}
\end{equation}

In canonical afterglow models that assume a uniform density environment, the 
cooling and injection breaks move to lower frequencies over time.  Eventually, 
both can lie below the absorption break, but far too late in the evolution of the 
burst to be relevant to anything but the radio afterglow, and long after the time 
at which the electrons in the jet have transitioned to the slow-cooling regime. In 
the more realistic density profiles in our models, the extremely high density that
the jet can encounter as it passes through a thick shell causes it to abruptly 
transition from highly relativistic to Newtonian expansion.  The decrease in 
$\Gamma$ leads to a sharp drop in the injection break frequency, while the 
increased density of the medium leads to a higher magnetic field amplitude, 
which in turn causes a drop in the cooling break frequency.  The result is that 
the frequency of the absorption break can be several orders of magnitude 
higher than the cooling and injection break frequencies as the jet passes 
through the shell.  Multiple transitions between the fast and slow electron 
cooling regimes can also occur. In a thick shell, when $\nu_a > \nu_m$ and the 
electrons are in the fast-cooling regime,

\begin{equation}
F_{\nu, \earth} = F_{\nu,\earth}^\text{max}
\begin{cases}
(\nu/\nu_c)^2(\nu_c/\nu_a)^{5/2}, &         \nu < \nu_c \\
(\nu/\nu_a)^{5/2},                & \nu_c < \nu < \nu_a \\
(\nu/\nu_a)^{-p/2},								& \nu_m < \nu.
\end{cases}
\label{eqn:fluxEquation3}
\end{equation}

\subsubsection{Slow-Cooling Electrons}

Our models yield the same flux as the canonical wind models until the jet collides 
with a shocked wind that has piled up behind a thick shell. If it does not reach the 
shocked wind in the first few hours, the electrons in the leading edge of the jet 
transition to the slow-cooling regime, with $\nu_a \ll \nu_m$ and

\begin{equation}
F_{\nu, \earth} = F_{\nu,\earth}^\text{max}
\begin{cases}
(\nu/\nu_a)^2(\nu_a/\nu_m)^{1/3},           &         \nu < \nu_a \\
(\nu/\nu_m)^{1/3},                          & \nu_a < \nu < \nu_m \\
(\nu/\nu_m)^{-(p-1)/2},                     & \nu_m < \nu < \nu_c \\
(\nu/\nu_c)^{-p/2}(\nu_c/\nu_m)^{-(p-1)/2}, & \nu_c < \nu.
\end{cases}
\end{equation}

\noindent If $\nu_m < \nu_a < \nu_c$,

\begin{equation}
F_{\nu, \earth} = F_{\nu,\earth}^\text{max}
\begin{cases}
(\nu/\nu_m)^2(\nu_m/\nu_a)^{5/2},           &         \nu < \nu_a \\
(\nu/\nu_a)^{5/2},                          & \nu_a < \nu < \nu_m \\
(\nu/\nu_a)^{-(p-1)/2},                     & \nu_m < \nu < \nu_c \\
(\nu/\nu_c)^{-p/2}(\nu_c/\nu_a)^{-(p-1)/2}, & \nu_c < \nu.
\end{cases}
\end{equation}

As noted above, as the jet passes through a dense shell, it can pass through 
multiple transitions between fast and slow electron cooling. When the electrons 
are in the slow-cooling regime and $\nu_a > \nu_c$,

\begin{equation}
F_{\nu, \earth} = F_{\nu,\earth}^\text{max}
\begin{cases}
(\nu/\nu_m)^2(\nu_m/\nu_a)^{5/2}, &         \nu < \nu_m \\
(\nu/\nu_a)^{5/2},                & \nu_m < \nu < \nu_a \\
(\nu/\nu_a)^{-p/2}, 							& \nu_a < \nu.
\end{cases}
\label{eqn:fluxEquation6}
\end{equation}

\subsection{Inverse Compton Scattering}

Low-frequency radiation can be boosted to higher energies in the presence of 
relativistic electrons through the process of inverse Compton (IC) scattering.  The 
contribution to the light curve at any frequency $\nu$ can easily be determined by
using the expressions for the synchrotron light curve at the IC frequency

\begin{equation}
\nu_\text{IC} = \frac{\nu}{\Gamma^2\left(1+\beta^2\right)}.
\end{equation}

\noindent The IC flux density at frequency $\nu$ is then

\begin{equation}
F_{\nu\text{, IC}} = \tau_e F_{\nu_\text{IC} \text{, syn}},
\end{equation}

\noindent where $F_{\nu_\text{IC} \text{, syn}}$ is the synchrotron flux density at 
frequency $\nu_\text{IC}$.

\subsection{Spherical Emission and Beaming}

The spherical nature of a GRB afterglow can substantially alter the observed light 
curve \citep{fenimoreEA96}.  The burst ejecta is initially ultra-relativistic, with $100 
\lesssim \Gamma \lesssim 1000$.  Radiation that is emitted by ejecta moving 
directly toward the observer (i.e., along the line connecting the observer and the 
progenitor) will be more highly beamed than radiation emitted by material that is 
not moving directly toward the observer.  There will also be a delay in the arrival 
time of photons that are emitted from regions of the jet not traveling directly toward 
the observer, as these regions are further away from the observer.  The beaming 
of radiation from material moving toward the observer leads to two effects.  First, a 
higher flux is observed than would be expected if beaming was ignored.  Second, 
the edge of the jet, which does not move directly toward the observer and is 
therefore not as highly beamed as the center of the jet, is not visible to the 
observer until $\Gamma \sim 1/\theta_j$.  The delay in arrival time of light emitted 
by material not moving directly toward the observer tends to smooth out sharp 
features in the light curve of the jet.  The method described above and in M12 
neglects the spherical nature of the emitting region but can be slightly modified 
to account for it.  The total luminosity $L'_{\nu'}$ radiated by the electrons at 
comoving frame frequency $\nu'$ must first be found.  As shown by \citet{
wijersGalama99}, the power per unit frequency transforms as $P_\nu = \Gamma P
'_{\nu'}$, implying that

\begin{equation}
L'_{\nu'} = 4\pi \frac{D^2 F_{\nu,\earth}}{\Gamma},
\end{equation}

\noindent where $D$ is the luminosity distance to the source and $F_{\nu}$ is 
the flux density detected at a distance $D$ from the source by an observer in 
the isotropic frame if the spherical nature of the emitting region is neglected.  
The observed flux density is then

\begin{equation}
F_{\nu} = \frac{1}{4\pi D^2} \iint\limits_{\Omega_j}{\frac{L'_{\nu'}[r(\theta)]\mathcal{D}
^3}{\Omega_j}d\cos{\theta}\ d\phi},
\label{eqn:correctedFluxEquation}
\end{equation}

\noindent where $\mathcal{D} = 1/{\Gamma(1-\beta\cos{\theta})}$ is the Doppler 
factor and $\Omega_j = 2\pi(1 - \cos{\theta_j})$ is the solid angle subtended by 
the jet \citep{moderskiEA00}.  Equation (\ref{eqn:correctedFluxEquation}) must 
be numerically integrated over the equal arrival time surface defined by 

\begin{equation}
t = \int{\frac{(1 - \beta\cos{\theta})}{c\beta}dr} = \text{constant} 
\label{eqn:equalArrivalTimesEqn}.
\end{equation}

Let $\left(r(t), \theta\right)$ refer to a point on the leading edge of the jet, with 
$0 \leq \theta \leq \theta_j$ and $\theta = 0$ referring to the center of the jet 
and the direction to the observer.  Radiation emitted by material at larger  
$\theta$ will be detected by the observer at progressively later times.  For a 
given observer time $t$, the jet radius $r$ at which radiation being detected 
from jet angle $\theta$ was emitted can be found by numerically integrating 
Equation (\ref{eqn:equalArrivalTimesEqn}) over $r$ with $\theta$ held 
constant until the desired value of $t$ is reached.  Equation 
(\ref{eqn:correctedFluxEquation}) is then used to find the total observed flux 
density by integrating over $\theta$ from $0$ to $\theta_j$.  Note that, due to 
the large cosmological redshifts that we study, another correction must be 
made to transform the isotropic frame flux density at a given frequency to the 
reference frame of an earthbound observer.

\end{document}